\documentclass{aa}  

\usepackage{natbib}
\usepackage{txfonts}
\usepackage[dvipsnames]{xcolor}
\usepackage{hyperref}
\hypersetup{
 colorlinks=true,
 citecolor=Violet,
 linkcolor=Red,
 urlcolor=Blue}
\usepackage[utf8]{inputenc}
\usepackage{mathrsfs}
\usepackage{graphicx}

\usepackage{soul} 
\begin{document} 
\title{TDCOSMO VIII: A key test of systematics in the hierarchical method of time-delay cosmography}

\author{Matthew R. Gomer\inst{1}\fnmsep \thanks{\email{mgomer@uliege.be}}
         \and
          Dominique Sluse\inst{1}
        \and
          Lyne Van de Vyvere\inst{1}
        \and
          Simon Birrer\inst{2,3,4}
        \and
          Frederic Courbin\inst{5}
        }

  \institute{STAR Institute, Quartier Agora - All\'ee du six Ao\^ut, 19c B-4000 Li\`ege, Belgium 
  \and 
  Kavli Institute for Particle Astrophysics and Cosmology and Department of Physics, Stanford University, Stanford, CA 94305, USA
  \and
  SLAC National Accelerator Laboratory, Menlo Park, CA, 94025
  \and
  Department of Physics and Astronomy, Stony Brook University, Stony Brook, NY 11794, USA
  \and
  Institute of Physics, Laboratory of Astrophysics, Ecole Polytechnique F\'ed\'erale de Lausanne (EPFL), Observatoire de Sauverny, 1290 Versoix, Switzerland
             }

  \date{Received date; accepted date}

 
  \abstract
    {The largest source of systematic errors in the time-delay cosmography method likely arises from the lens model mass distribution, where an inaccurate choice of model could in principle bias the value of $H_0$. A Bayesian hierarchical framework has been proposed which combines lens systems with kinematic data, constraining the mass profile shape at a population level. The framework has been previously validated using a small sample of lensing galaxies drawn from hydro-simulations. The goal of this work is to expand the validation to a more general set of lenses consistent with observed systems, as well as confirm the capacity of the method to combine two lens populations: one which has time delay information and one which lacks time delays and has systematically different image radii.
    For this purpose, we generated samples of analytic lens mass distributions made of baryons+dark matter and fit the subsequent mock images with standard power-law models. Corresponding kinematics data were also emulated. The hierarchical framework applied to an ensemble of time-delay lenses allowed us to correct the $H_0$ bias associated with model choice to find $H_0$ within $1.5\sigma$ of the fiducial value. We then combined this set with a sample of corresponding lens systems which have no time delays and have a source at lower $z$, resulting in a systematically smaller image radius relative to their effective radius. The hierarchical framework has successfully accounted for this effect, recovering a value of $H_0$ which is both more precise ($\sigma\sim2\%$) and more accurate ($0.7\%$ median offset) than the time-delay set alone. This result confirms that non-time-delay lenses can nonetheless contribute valuable constraining power to the determination of $H_0$ via their kinematic constraints, assuming they come from the same global population as the time-delay set.
    }

    \titlerunning{Hierarchical time-delay cosmography systematics}
    \authorrunning{M. R. Gomer et al.}
    \maketitle
%

\section{Introduction}
A precise determination of the Hubble constant $H_0$ is a widely sought after goal, now with multiple mature measurement methods. Some methods measure the early universe and recover $H_0$  \citep[e.g.,][]{Aiola20,Planck20,Abbott18}, while some use standard candles to measure $H_0$ locally \citep[e.g.,][]{Riess22,Freedman20}. There is a discrepancy between these measurements which may require the addition of new physics into the standard cosmological model \citep[e.g.,][]{Poulin19}, as extensive  considerations of possible unaccounted-for systematic effects in one or more methods of $H_0$ determination have not provided an explanation yet \citep[e.g.,][]{Freedman21,Huang20,DArcy19,Addison18}. Additional independent measurement methods may help diagnose the tension, and should be explored before making claims of new physics. Many such novel methods have been developed, such as using water megamasers \citep{Pesce20} or gravitational sirens \citep{Abbott17} to directly measure distance. One such distance-ladder-independent method is the gravitational lensing time-delay method.

The gravitational lensing method measures distances through the use of time delays between different images of the same variable source. The combination of the observed time delays with the lens mass model gives a time-delay distance:
\begin{equation}
D_{\Delta t} = (1+z_d) \frac{D_d D_s}{D_{ds}}\propto H_0^{-1},
\end{equation}
where $z_d$ is the redshift of the lens (deflector), and $D_d$, $D_s$, and $D_{ds}$ are the angular diameter distances from the observer to the lens, observer to the source, and deflector to the source, respectively. Distance (i.e., $H_0$) measurements can be combined between numerous systems to increase the statistical precision. The method requires high-quality imaging data, long-term time-delay monitoring, and precise spectroscopy for kinematic measurements. Large collaborations leading the efforts with this method include COSMOGRAIL  \citep{Courbin18,Bonvin19,Millon20}, SHARP \citep{Spingola18,Chen19}, STRIDES \citep{Treu18,Anguita18,Shajib19a,Shajib20}, and H0LiCOW  \citep{Suyu17,Wong20,Rusu20}, which together form the TDCOSMO umbrella collaboration \citep{TDCOSMO1}.

The primary challenges for the lensing method are degeneracies, whereby the same lensing observables (e.g., image positions and relative fluxes) are recovered from a set of mass distributions. Lensing information alone has no ability to constrain which distribution within the set is the truth. Of particular interest is the mass sheet degeneracy \citep[MSD,][]{Falco85}, a specific case of the more general source position transformation \citep[SPT,][]{Schneider13,Schneider14,Unruh17,Wertz18}. The MSD is extensively studied and well-documented in the literature \citep[e.g.,][]{Saha06, Xu16, Birrer16, Sonnenfeld18, TDCOSMO4, Kochanek20,Kochanek21, Gomer20, Blum20, TDCOSMO8, Birrer21b}. The MSD is described through the mass sheet transformation (MST), in which a mass distribution $\kappa(\theta)$ can be scaled by a factor of $\lambda$ with the addition of a corresponding constant, analogous with the addition of a uniform sheet of mass, into any one realization from a set of transformed distributions:
\begin{equation}\label{eq:mst}
    \kappa_{\lambda}(\theta) = \lambda \kappa(\theta) + (1-\lambda),
\end{equation}
where we have used the $\lambda$ subscript to indicate the quantity has been transformed by an MST. The physical nature of a mass sheet is often conceptually separated into two different phenomena: an "external" mass sheet coming from additional mass along the line of sight, and an "internal" mass sheet which encodes the possibility of an alternative lens profile shape. This work concerns the internal mass sheet. Together with a corresponding transformation of the unobserved source position $\beta_\lambda = \lambda \beta$, the MST leaves image positions and relative fluxes unaffected, while leading to a multiplicative constant on the time-delay distance measurement $D_{\Delta {t}\lambda}=\lambda^{-1}D_{\Delta {t}}$. Since the recovered value of $H_0$ is inversely proportional to the time-delay distance, a determination of $H_0$ from a mass-sheet-transformed model results in a biased value of $H_{0\lambda}=\lambda H_0$. As such, the capacity of lens modeling to recover an accurate measure of $H_0$ is inherently a problem of constraining $\lambda$. 

Since lensing information alone is unable to constrain $\lambda$, one must turn to external information. The impact of the MST can be mitigated via stellar velocity dispersion measurements, which provide an independent measurement of mass \citep{Treu02a,Treu02b,Koopmans03, Barnabe07}. This addition requires that a given mass distribution reproduces both the lensing data and kinematic data, which places constraints on the profile shape, and therefore the range of allowable $\lambda$.

To further constrain $\lambda$, it is common to incorporate existing knowledge about galaxy structure through the implementation of a parametric profile. Early-type galaxy mass profiles are known to be well-approximated as a power law with respect to radius with approximately isothermal slope, a phenomenon sometimes referred to as the bulge-halo conspiracy \citep{Gavazzi07, Suyu09, Koopmans09, Humphrey10, Auger10, Barnabe11, Dutton14}. By specifically using a power law to fit a lens, a modeler fixes $\lambda$ by restricting the fit to the $\kappa_{\lambda}$ which matches the power-law profile form over the relevant radii. As such, the choice of a specific family of model, power-law or otherwise, restricts the range of possible MSTs which fit the data. If a modeler uses a power law to fit a lens which in reality differs from a power law, the MST which causes the profile to best match a power law is implicitly applied, which can result in a biased value of $H_0$. One approach to account for this effect is to use multiple different families of models for the same lens, which is why H0LiCOW uses both a power-law and a two-component composite model to fit time-delay lenses for cosmography \citep{Suyu14, Rusu20}. TDCOSMO's exploration of systematics found agreement in $H_0$ when using a power law or composite model in the fit, providing supporting evidence that this MST effect is not severe in real lenses \citep{TDCOSMO1}. When using a power-law model for 6 lenses supplemented with aperture kinematic measurements, H0LiCOW found $H_0 = 73.3_{-1.8}^{+1.7} $ km s$^{-1}$ Mpc$^{-1}$ \citep{Wong20}.


\citet{TDCOSMO4} (hereafter TDC4) introduced a technique to supplement the parametric method
by explicitly folding the effect of the MSD into a hierarchical analysis of the lens systems. This method is able to combine sets of lens modeling results together, provided that the sets of lens galaxies come from the same overall galaxy population. The method applies an MST to the lens modeling results on a population level, serving to capture systemic departure from the initial parametric model. To account for the difference of lensing impact parameter arising from having different lens and source redshifts, the amplitude of the MST is allowed to vary as a function of the ratio between the effective radius and Einstein radius. By directly parameterizing and including this effect, this approach allows the velocity dispersion data of the collective set to constrain the space of possible MSTs.
Specifically, systems from the Sloan Lens ACS Survey \citep[SLACS,][]{Bolton06, Bolton08, Shajib21}, which have lens models and velocity dispersions, but lack time delays, are able to be included in the analysis to place constraints on the population-scale MST, given the hypothesis holds that this population comes from the same global population as that of the time-delay lenses. We detail this method in the following section.

The hierarchical framework has been validated for a population of time-delay lenses by TDC4 based on the simulated lenses from Rung 3 of the Time Delay Lens Modeling Challenge (TDLMC) \citep{Ding18,Ding21b}. The value of $H_0$ was inferred by modeling hydrodynamically simulated lensed systems where the lensing galaxy was drawn from a set of 16 massive elliptical galaxies from Illustris-TNG and from zoom-in simulations. The resulting value of $H_0$ ($68.4_{-3.7}^{+3.4}$ km s$^{-1}$ Mpc$^{-1}$ when using a uniform prior for anisotropy radius) was consistent with the fiducial input value ($65.413$ km s$^{-1}$ Mpc$^{-1}$) with most of the uncertainty coming from limited constraints on the anisotropy of the velocity dispersion. However, the TDLMC Rung 3 lenses are known to have numerical effects in the central region, for which the authors caution against their strict interpretation as realistic lenses \citep{Ding21b}. We also note that the TDLMC comparison does not test the inclusion of a corresponding non-time-delay lens population. With this in mind, the efficacy of the hierarchical method for real systems is supported, but not confirmed. A test of the method using analytically created mock systems which do not have such artificial cores would supplement this result, which is one goal of this work.

Combining the seven TDCOSMO systems, TDC4 recovered a value of $H_0 = 74.5_{-6.1}^{+5.6}$ km s$^{-1}$ Mpc$^{-1}$, which includes the MSD in the error budget. This is consistent with the \citet{Wong20} value, but with widened uncertainty. Seeking better constraints, 33 non-time-delay lenses from SLACS \citep{Shajib21} were implemented into the hierarchical framework, under the assumption that the SLACS and TDCOSMO lenses are part of the same population. Interestingly, the resulting value of $H_0$ moves downward to $67.4_{-3.2}^{+4.1}$ km s$^{-1}$ Mpc$^{-1}$. While these two values are statistically consistent, the hierarchical method nevertheless relies on different assumptions than other methods and so it must be further tested, motivating the experiment in this paper.

The main concern is that if there were some systematic difference between the SLACS and TDCOSMO lens populations, the primary assumption of the hierarchical framework would be flawed, and the result would be unreliable. Unfortunately, we do not have access to the true mass distributions of these galaxies, and as such we cannot confirm or reject the possibility that both samples are drawn from the same population. We can, however, perform targeted tests of systematics; a natural place to begin comes from the fact that the SLACS and TDCOSMO lenses have different redshifts, and therefore probe different radii relative to the effective radius. 
Ergo, the second goal of this paper is to test the capacity of the hierarchical framework to correctly capture and account for this change in local lensing effect which stems from a difference in redshift. This goal can be addressed by by creating mock lenses consistent with the SLACS and TDCOSMO lens populations and combining them together hierarchically.


In summary, in this work we further test the hierarchical framework by modeling a large sample of mock lensed systems meant to represent those of the TDCOSMO galaxies, constructed as analytical two-component light+dark matter mass distributions. 
Specifically, we test the capacity of the framework in two respects. Firstly, we wish to confirm that the framework still performs well with intrinsic mass distributions which are less cored in their center than galaxies from hydro-simulations. Secondly, we wish to confirm that the inclusion of the comparison sample of non-time-delay lens systems is correctly accounted for without introducing a bias arising from the change in redshift (i.e., lensing impact parameter).
To address this second respect, we construct our lens population such that it is consistent with the TDCOSMO population at one source redshift, and consistent with the SLACS population when the source redshift is changed.


We first describe the hierarchical framework and experiment setup in more detail in Sect. \ref{sec:hierarc}, then outline our mock creation in Sect. \ref{sec:mock_creation}, present our results in Sect. \ref{sec:hierarc_results} and discuss these results in Sect. \ref{sec:discussion}. 

\section{Hierarchical framework and experiment setup} \label{sec:hierarc}
The hierarchical framework is detailed in TDC4, but we summarize the main components here. The key idea of the hierarchical framework is that information on the amplitude of the MST can be retrieved by considering a population of lens systems with kinematic data. While single-aperture kinematics can place limited constraints on $\lambda$ for individual systems, the framework is able to leverage the additional statistical constraining power of a population of systems.

To gain a physical intuition of the method, consider that when an individual lens is fit as a power law (PL), the actual constraint is that the lens profile is a PL+MST with an unknown $\lambda$. However, these galaxies as a population are thought to follow approximately the same global mass profile (with a characteristic radius that scales with the luminous effective radius, $\theta_{\rm eff}$\footnote{
    Our notation is to use $\theta$ to refer to a 2D projected radius in angular units, $R$ to refer to the same quantity in physical kpc, and $r$ to refer to the unprojected 3D radius.}),
meaning the $\lambda$ necessary to match a given lens to a PL is purely a function of which part of the profile the Einstein radius ($\theta_{\rm Ein}$) lands upon.
Kinematics also probe some radial range of this global profile, depending on the size of the measurement aperture relative to $\theta_{\rm eff}$, but unlike lensing information, kinematics are sensitive to $\lambda$, as they measure the true mass within a spherical shell. As such, the combination of many systems together can be thought of as probing the global profile at a range of radii with both lensing and kinematic data, where the measurements from one system can help support the unconstrained parts of another system. This combined information allows one to infer the population-scale $\lambda$ behavior associated with the mapping of the global profile into a PL.

One can show that the MST-invariant term extracted from a lensing measurement is:
\begin{equation}\label{eq:xidef}
    \xi=\theta_{\rm Ein}\frac{\alpha''_{\rm E}}{1-\kappa_{\rm E}},
\end{equation}
where $\alpha''_{\rm E}$ is the circularly averaged radial second derivative of the deflection evaluated at $\theta_{\rm Ein}$, and $\kappa_{\rm E}$ is the mean convergence of an annulus at $\theta_{\rm Ein}$ \citep{Sonnenfeld18,Kochanek20}. From this point of view, the hierarchical inference takes a set of $\xi$ constraints at different Einstein radii and uses the kinematic information to provide the conversion from the MST-agnostic description into the true mass distribution with an associated $\lambda$, which is allowed to be different for different Einstein radii. 

The inference works by ascribing a distribution of MSTs to the galaxy population, where the $\lambda_{\rm int}$ distribution is described with Gaussian standard deviation $\sigma(\lambda_{\rm int})$. The "int" denotation references the conceptual separation between an internal mass sheet transformation as described here from an external mass sheet, owing to convergence along the line of sight $\kappa_{\rm ext}$. To account for differences in the radii at which the global mass profile is probed, a linear scaling  with the effective radius relative to the Einstein radius is included as 
\begin{equation} \label{eq:mst_parameterization}
    \lambda_{\rm int}(\theta_{\rm eff}/\theta_{\rm Ein})=\lambda_{\rm  int,0}+\alpha_{\lambda} \left( \frac{\theta_{\rm eff}}{\theta_{\rm Ein}}-1\right)
\end{equation}
with slope $\alpha_{\lambda}$ and intercept $\lambda_{\rm int,0}$. We note the distinction between the set of $\lambda_{\rm int}$ values each lens within a population individually undergoes and the intercept of our linear parameterization of the population, $\lambda_{\rm int,0}$, a descriptive parameter we recover. This scaling allows for lenses which are probed at different radii to have different MST amplitudes to match the power-law shape. The constraint on $\lambda_{\rm int}$ comes from kinematic information, for which even lenses without time delays can provide meaningful constraints. Lens population kinematics are included in a spherical Jeans framework parameterizing the distribution of anisotropy $a_{\rm ani}$ (further detailed in Sect. \ref{ssec:kinematics}) as Gaussian with mean $\langle a_{\rm ani}\rangle$ and standard deviation $\sigma(a_{\rm ani})$. In total, the hierarchical framework recovers values for $\lambda_{\rm int,0}$, $\sigma(\lambda_{\rm int})$, $\alpha_{\lambda}$, $\langle a_{\rm ani}\rangle$ and $\sigma(a_{\rm ani})$ to describe the distribution of values within the population and place an informed constraint on $H_0$.

The hierarchical inference has been successfully tested on numerically simulated TDLMC profiles, which have artificial cores arising from numerical effects. However, it has not yet been confirmed to be effective for analytical profiles which do not have such cores. In addition to testing this, we sought to test the capacity of the method to incorporate an external non-time-delay lens population which is similar to the time-delay lens population, except with a different relative impact parameter, (i.e., different $\theta_{\rm eff}/\theta_{\rm Ein}$), as was done for real systems in TDC4.

To control for all other variables, we chose for the two lens populations to be identical, except for the source redshift, for which we selected one value for each population. Given the same lens surface mass density $\Sigma(\theta)$, a larger source redshift $z_s$ decreases the critical lensing density $\Sigma_{\rm crit}$ and hence increase the convergence $\kappa(\theta)$:
\begin{equation} \label{eq:sigcrit}
    \centering
    \Sigma_{\rm crit}=\frac{c^2}{4\pi G}\frac{D_s}{D_d D_{ds}}, \qquad \kappa(\theta,z)=\frac{\Sigma(\theta)}{\Sigma_{\rm crit}(z_d, z_s)}.
\end{equation}
This increased $\kappa$ leads to a larger Einstein radius, changing which part of the profile is probed by lensing. If a profile has a changing shape with radius, probing different radii could change the resulting fit, resulting in a different MST and ultimately in a change in $H_0$. As such, the expected change in $\lambda_{\rm int}$ with respect to impact parameter encodes two effects: the change in redshift leading to a different probed radius of a given profile, and the variety of profile shapes arising from the set of two-component mass distributions at a given redshift. The hierarchical setup assumes that these effects collectively are well-described as linear with Gaussian scatter (Eq. \ref{eq:mst_parameterization}). If the combined effect is not well-described by this parameterization, there will be nonrandom departures from the linear description, resulting in an incorrect scatter estimation and a biased $H_0$. However if the linear parameterization is a good description of the combined effect, such that the variations between systems can be captured by the Gaussian scatter, the effect is accounted for and the correct $H_0$ will be recovered. As such, an accurate $H_0$ recovery in this experiment represents supporting evidence for the versatility of the hierarchical method. On the other hand, if the method were to fail in the case where the only difference between populations comes from source redshifts, the hierarchical framework in its current form would be unlikely to be reliable for the more complicated real population. 

This experiment is not necessarily intended to perfectly emulate the data sets for either SLACS or TDCOSMO specifically. Instead, the intention is to perform a targeted test of the effect of probed radius within the hierarchical framework, using a set of lenses engineered to come from the same population. 

\section{Creating mocks} \label{sec:mock_creation}
We constructed a set of two-component profiles consistent with the observed lens population. For each profile, we simulated two mock HST-quality images corresponding to the lens with two different source redshifts. Each lens image was modeled using a power-law profile for the lensing galaxy. Mock kinematic measurements were calculated and combined with the lensing posteriors in a hierarchical framework. This section details how these mass profiles were created (\S \ref{ssec:mass_distribution}), how the population of these profiles compares to the observed lenses (\S \ref{ssec:lenspop}), how mock images were created and fit (\S \ref{ssec:mockimages}), and how kinematics were simulated (\S \ref{ssec:kinematics}).

\subsection{Mass distribution}\label{ssec:mass_distribution}
We constructed lenses using a two-component composite mass model consisting of a Chameleon  profile \citep{Dutton11,Suyu14} representing the light distribution and an NFW \citep{Navarro96, Navarro97} component representing the dark matter. We assumed that the mass distribution of the baryon component corresponds to the light distribution of observed systems. Most observed light profiles of early-type galaxies are modeled as S\'ersic profiles, for which the corresponding surface mass density is given as
\begin{equation}\label{eq:sersic_prof}
    \Sigma(\theta_x,\theta_y)=\Sigma_{\rm eff} \  {\rm exp}\left\{-b_n\left[\left(\frac{\sqrt{q \theta_x^2+\theta_y^2/q}}{\theta_{\rm eff}}\right)^{1/n}-1\right]\right\},
\end{equation}
with axis ratio $q$ where $\theta_{\rm eff}$ is the effective radius enclosing half of the total light and $n$ is the S\'ersic index \citep{Sersic63,Ciotti99}. The constant $b_n$ is set to ensure the half-light property of $\theta_{\rm eff}$ given $n$. Unfortunately the elliptical S\'ersic profile requires numerical integrals in the lens deflection calculation and as such is computationally expensive in this context. We therefore chose to represent the baryonic matter as a Chameleon profile \citep{Dutton11}, which is a profile designed to closely match a S\'ersic profile at lensing radii using a combination of two softened isothermal profiles \citep{Kassiola93},
\begin{equation}\label{eq:cham_prof}
\begin{aligned}
    \Sigma(\theta_x,\theta_y)=\frac{\Sigma_0}{(1+q)}\left[\frac{1}{\sqrt{q \theta_x^2+\theta_y^2/q+4w_c^2/(1+q)^2}}\right. \\
   \left. -\frac{1}{\sqrt{q \theta_x^2+\theta_y^2/q+4w_t^2/(1+q)^2}}\right]
\end{aligned}
\end{equation}
with parameters $w_c$ and $w_t$ with $w_t>w_c$.  The Chameleon profile is often used for this purpose in the context of lensing \citep[e.g.,][]{Suyu14,Birrer19, Rusu20, Shajib20, TDCOSMO7}. 

Given a S\'ersic index and $\theta_{\rm eff}$, one can use the prescription of \citet{Dutton11} to construct an analogous Chameleon profile. We note that this prescription claims that the Chameleon profile matches the enclosed mass of the S\'ersic profile to within a few percent within the $[0.5\theta_{\rm Ein},3\theta_{\rm Ein}]$ range, but actually requires a modification to reach this reported precision, which we have implemented and discussed in Appendix \ref{sec:cham_troub}. With this, we were able to construct a Chameleon profile which closely matches a provided S\'ersic. 

The NFW profile is described in 3D mass density as 
\begin{equation}\label{eq:nfw_prof}
    \rho(r)=\frac{\rho_0}{(r/r_s)(1+r/r_s)^2}
\end{equation} with scale radius $r_s$ and scale density $\rho_0$.  Deflection angles and lens potentials for an elliptical NFW in projection can be calculated. Following \citet{Golse02}, ellipticity is added in the lens potential.
The parameters for these Chameleon+NFW profiles must be selected to result in a population which is consistent with the observed SLACS and TDCOSMO lensing galaxies  \citep{Dutton14, Shajib21}.

\subsection{Creating a lens population} \label{ssec:lenspop}
\begin{figure*}
    \centering
    \includegraphics[width=\linewidth]{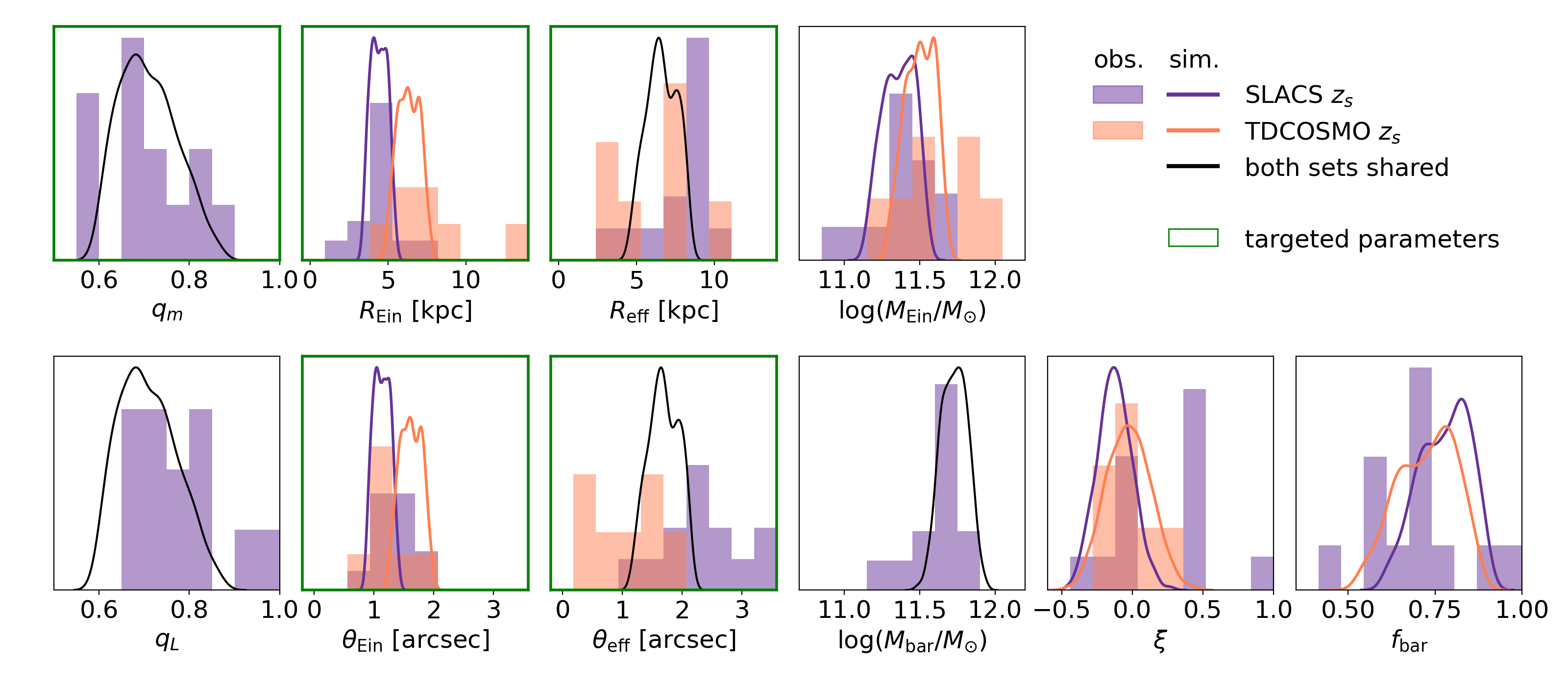} 
    \caption{Histograms depict the distribution of observed SLACS (purple) and TDCOSMO (orange) lens properties. Our simulated populations (shown as KDEs of corresponding colors) were created by constructing a large number of profiles and selecting a subset which simultaneously match the 5 target parameter distributions, indicated by the green borders. The simulated SLACS and TDCOSMO populations are the same profiles, only differing by a changed source redshift. For parameters where the two sets are identical, a black line is used instead of orange or purple.
    }
    \label{fig:param_properties}
\end{figure*}

To build up a realistic population of lens systems, we examined the existing population of lens galaxies which have been modeled as a composite profile. We considered the subset of 14 SLACS lenses used by TDC4 which have been modeled by \citet{Shajib21} and therefore have fit constraints on their slope, $\gamma$. These lenses also have stellar masses measured via stellar population analysis using a Salpeter IMF \citep{Auger09}. Combining these measurements, we examined the distribution of several observed properties (i.e., axis ratio of the light $q_L$, $\theta_{\rm eff}$, $\theta_{\rm Ein}$) as well as some recovered properties: axis ratio of the mass ($q_m$), mass within $\theta_{\rm Ein}$ ($M_{\rm Ein}$), total mass in baryons ($M_{\rm bar}$), baryon fraction within $\theta_{\rm Ein}$ ($f_{\rm bar}$), and the MST-invariant $\xi=2(\gamma-2)$ for a power law. We chose to plot $\xi$ as recovered from the lensing power-law fits rather than $\gamma$ directly to better reflect that the fit slope is a property recovered from lensing and may not correspond to the true local slope at the Einstein radius \citep{Kochanek20}. Additionally, we compiled the same properties for the TDCOSMO set of 7 lenses, although we lack measurements for $q$, $M_{\rm bar}$ or $f_{\rm bar}$, and plotted the two populations as histograms in Fig. \ref{fig:param_properties}. 

We sought to create a synthetic population of two-component profiles which closely matches both observed populations.
To create a two-component profile, 7 parameters (omitting centroid positions and position angles) had to be given values. For the baryon component we have $\Sigma_{\rm eff}$, $\theta_{\rm eff}$, $n$, and $q_L$. For the dark matter halo component we have $\rho_0$, $r_s$ and $q_h$. Furthermore, we also must select redshifts for the source of each lens.

The first simplifying assumption we made is that the two components have the same axis ratios ($q=q_L=q_h$). We also set the position angles and centroids to be aligned. In reality, these components need not be aligned \citep{Gomer18,Shajib19a}, and have been allowed to vary in lensing models \citep[e.g.,][]{Rusu20,TDCOSMO1}. However, since the primary focus for this work is the effect of the change in Einstein radius regarding the radial profile, rather than the effects of the angular structure of the lens, we fixed the components to be aligned. This assumption is supported by the joint lensing+dynamics analysis of \citet{Shajib21}, who find that centroids of light and mass match within 218\,pc (68\% confidence) and position angles  align within $10\deg$ for most SLACS galaxies, consistent with earlier results for quasar lenses \citep{Keeton98,Yoo06,Sluse12}. Numerical results using the Illustris simulation also support this level of alignment \citep{Xu17}. We therefore chose to adopt this simple angular structure for the input mocks, although the lens model we use to fit the systems does have additional azimuthal freedom through external shear.

\begin{figure}
    \centering
    \includegraphics[width=0.98\linewidth]{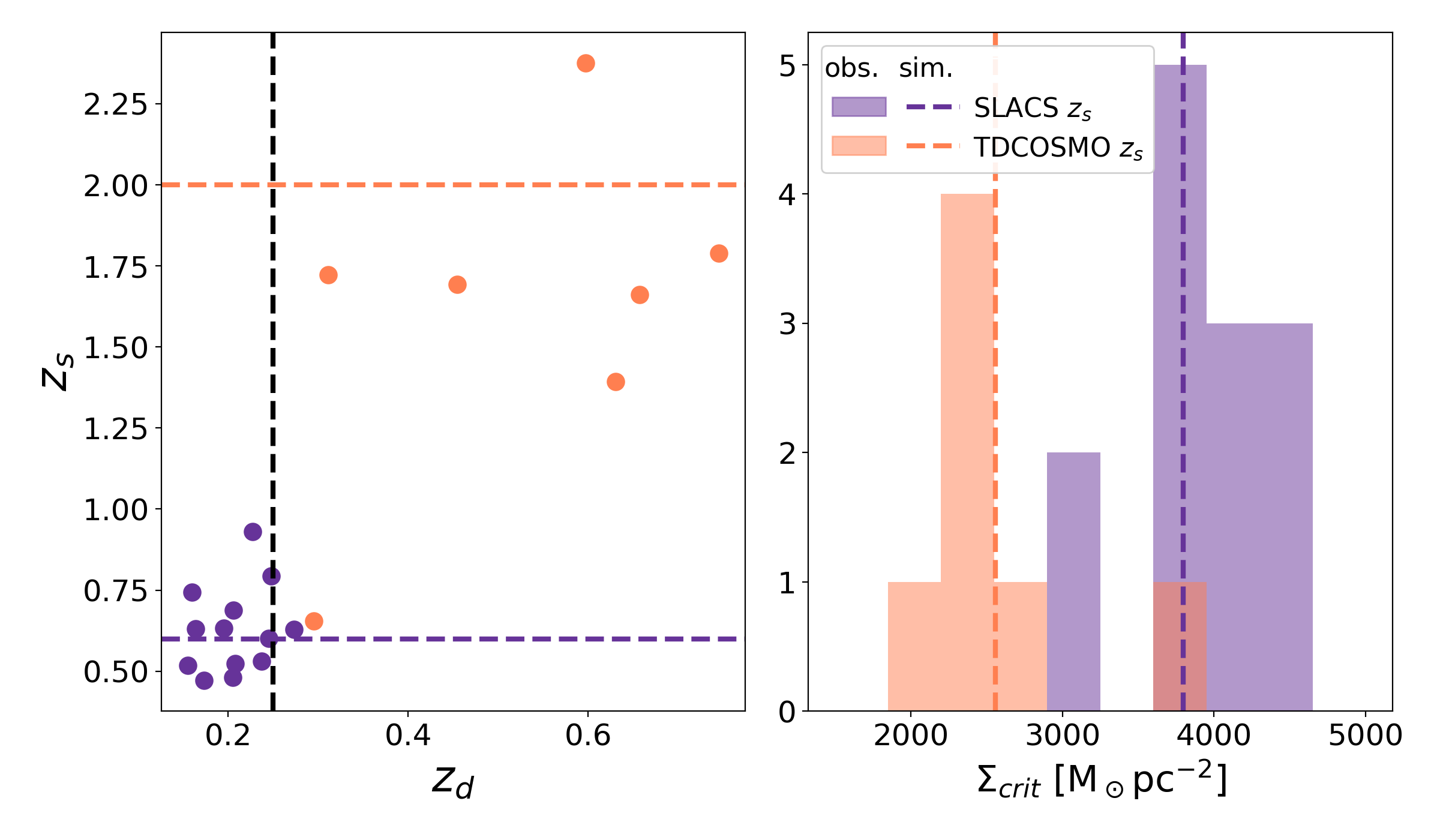} 
    \caption{Redshift comparison with the observed lens population. Left: the distributions of source redshifts against lens redshifts for the populations of observed SLACS (purple) and TDCOSMO (orange) lenses. Our selections of source and lens redshifts are indicated as dashed lines. Right: these selections result in the dashed-line $\Sigma_{\rm crit}$ values, which are consistent with the observed populations.
    }
    \label{fig:redshiftdist}
\end{figure}

The intention of our mock population is that it resemble the SLACS distributions for a particular source redshift, $z_s$, and the TDCOSMO distributions after a change to another $z_s$. In reality, both populations exist over a range of $z_s$ and lens redshift $z_d$, shown in Fig. \ref{fig:redshiftdist}. Generally, the TDCOSMO set has larger $z_s$ and $z_d$. For the targeted experiment in this work, we selected just one $z_d$ to be the same for both sets to set aside the effects of a changing angular scale (which affects image resolution and kinematic aperture size) and a changing cosmological scale factor (which changes comoving values of effective radii). We selected $z_d=0.25$, which is higher than the typical SLACS value and lower than the typical TDCOSMO value. Strictly speaking, this lens redshift is below the lowest TDCOSMO $z_d$, but serves as a nice middle ground which is capable of retaining a $\Sigma_{\rm crit}$ for each set which is consistent with each respective population. We selected just two $z_s$: one to represent each of our two populations. 
We selected a $z_s$ for what we call the "SLACS-like" set to be 0.6 and a $z_s$ for what we call the "TDCOSMO-like" set to be 2.0. These redshift selections result in $\Sigma_{\rm crit}$ values which are close to the medians of those of the respective SLACS and TDCOSMO populations. These calculations use a flat universe with $H_0=70$  km s$^{-1}$ Mpc$^{-1}$ and $\Omega_M=0.3$.

For the baryonic mass component, we set the S\'ersic index of the light component to be $n=4$, equivalent to a de Vaucouleurs profile \citep{Sersic63}. We then used the prescription described in Appendix \ref{sec:cham_troub} to create a Chameleon profile from an input $\Sigma_{\rm eff}$ and $\theta_{\rm eff}$. 

Even once a model family is chosen, there still is considerable freedom in the handful of parameters required to make a lens. Some parameters (such as $q$) can be directly observed and inserted into a model, while others (such as NFW scale radius $r_s$) do not easily convert into the corresponding observable constraint (like $\theta_{\rm Ein}$, which indirectly is determined by $r_s$, in conjunction with other parameters). We wish to correctly probe the space of observed values, but must create our mocks using parameter values which are sometimes not observable. To solve this problem, we randomly sampled the input parameters over a range, resulting in a large set of profiles with would-be observed values. Comparison to the observed distribution provides a set of target parameters, from which one can select only the realizations which match the targets to our desired precision. This was done by defining a $\chi^2$ for each targeted quantity and selecting a subset of profiles below a threshold total $\chi^2$. We selected a population which targets the observable quantities $q$, $\theta_{\rm Ein}$, and $\theta_{\rm eff}$.\footnote{
    Since we targeted each radius both in terms of physical size and angular size, the effective radius and Einstein radius were effectively weighted twice the weight of $q$.}
    \footnote{
         The Chameleon profile drops more quickly at large radii than the S\'ersic it is meant to emulate. As such, the mass at infinity of the baryon component may not perfectly match a S\'ersic. For $\theta_{\rm eff}$ (and $R_{\rm eff}$), the value of integrated light at infinity of the S\'ersic profile was used, rather than the artificially truncated Chameleon. The other quantities which are not defined using behavior at infinity are expected to be sufficiently accurate: the Einstein radius changes by about 0.01\arcsec comparing a S\'ersic+NFW to a Chameleon+NFW. As such, for calculation of the quantities other than $\theta_{\rm eff}$ (and $R_{\rm eff}$), the Chameleon profile was used directly.}

\begin{figure}
    \centering
    \includegraphics[width=0.85\linewidth]{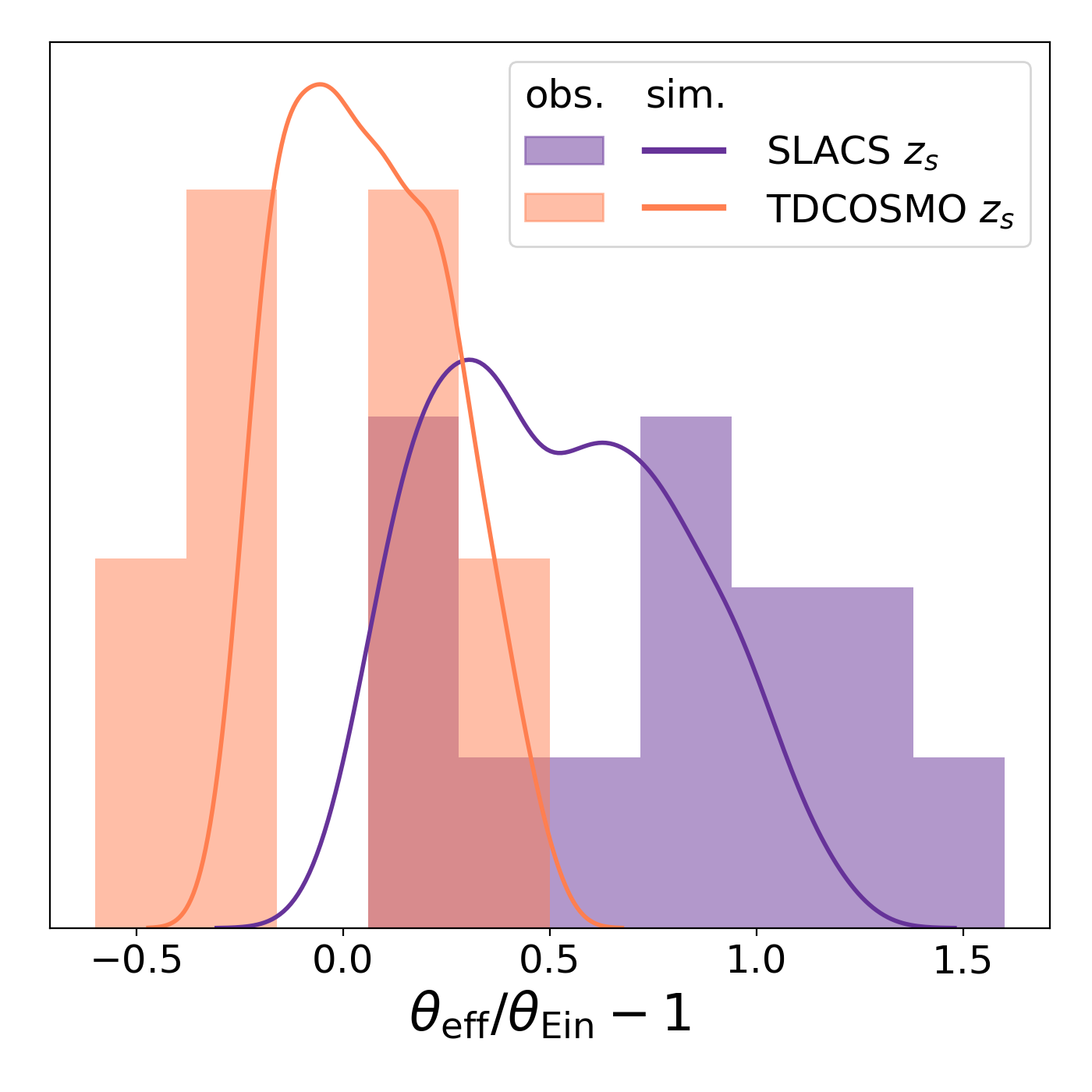} 
    \caption{Scaled radial quantity used by the hierarchical framework to parameterize radial dependence for MST, plotted for the observed population (histograms) and the simulated profiles used in this work (KDE curves). 
    }
    \label{fig:scaling_param}
\end{figure}

After selecting a number of profiles which met our targets, we arrived at a population of 465 lens profiles, which are a good match to the observed lens populations\footnote{
        The only minor difference arises because of the $z_d=0.25$ compromise, which resulted in the physical scale conversion being different for the TDCOSMO lenses than for our lenses, and so even though the Einstein radius and effective radius match well in terms of kpc, our synthetic values are slightly larger than the median TDCOSMO values in terms of arcseconds.},
with distributions of parameter values approximated via Kernel Density Estimation (KDE) in Fig. \ref{fig:param_properties}. Aside from the intended match to the target parameters, we were also able to compare some nonobservable quantities which were not targeted directly, such as baryon fraction or total mass. These quantities also match well between the observed and simulated populations, with the possible exception that the recovered values of $\xi$ were closer to isothermal than some of the observed SLACS lenses, which recover steeper slopes (recall that $\xi=0$ corresponds to an isothermal slope).
Nevertheless, our simulated distribution overlaps significantly with the real lenses. In particular we have reproduced the difference in impact parameter between SLACS and TDCOSMO lenses: Fig. \ref{fig:scaling_param} shows that TDCOSMO-like lenses have images forming at a larger distance (smaller $\theta_{\rm eff}/\theta_{\rm Ein}$) than SLACS-like systems. This agreement with the observed data is important for testing the relevance of the parameterization of the MST with radius (Eq. \ref{eq:mst_parameterization}) used in the hierarchical framework.

\begin{figure}
    \centering
    \includegraphics[width=\linewidth]{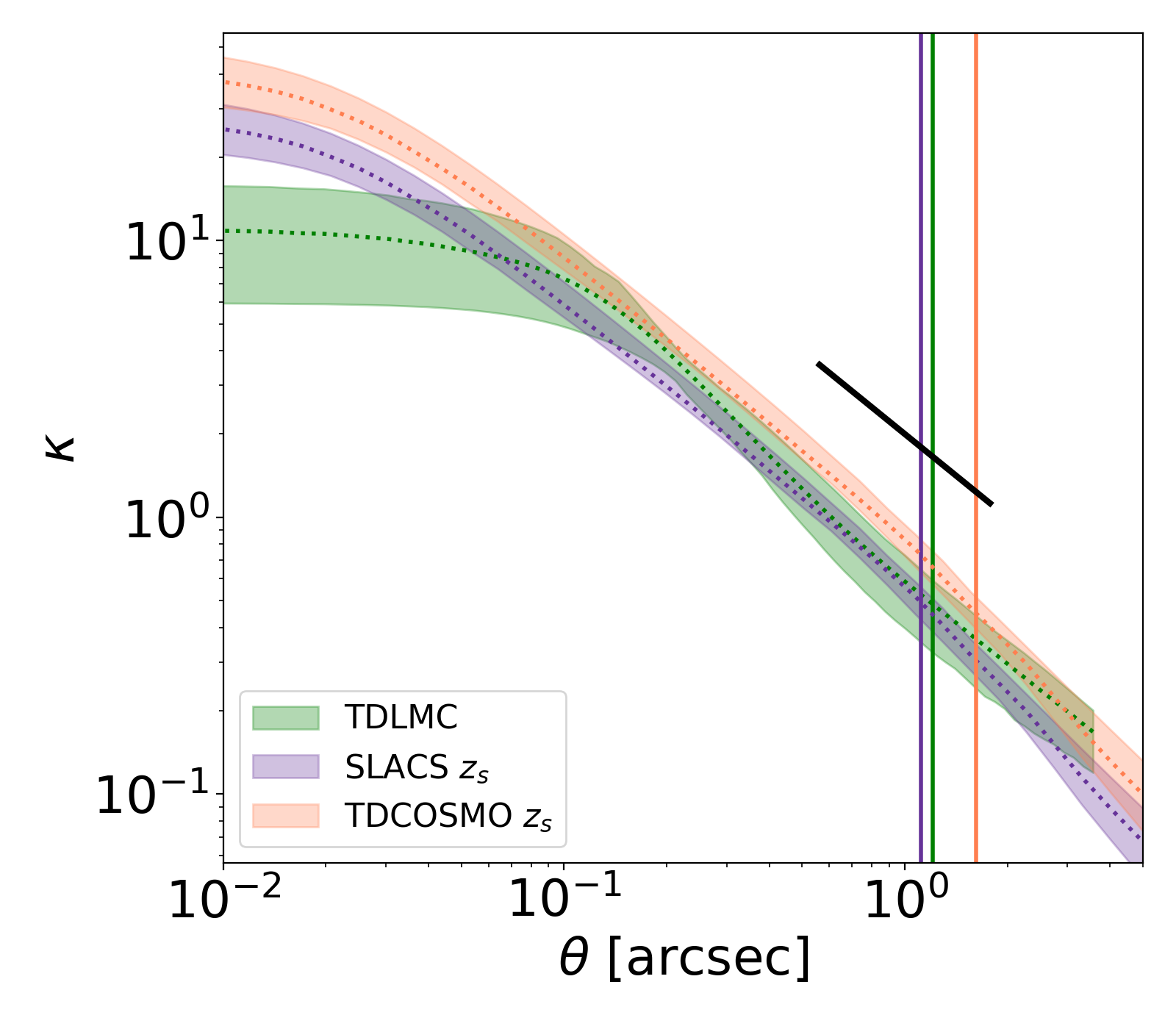} 
    \caption{Convergence profiles for 20 analytical Chameleon+NFW profiles used in this paper (purple: SLACS-like; orange: TDCOSMO-like) alongside 16 profiles from hydro-simulations used in the TDLMC (green). Colored bands correspond to the 16th and 84th percentiles. Vertical colored lines indicate the median Einstein radius of the corresponding population. The diagonal black line indicates an isothermal slope for comparison.
    }
    \label{fig:kappa_profiles}
\end{figure}

We plot the convergence profiles as a function of radius for our mock populations in Fig. \ref{fig:kappa_profiles}. In comparison with the hydrodynamically simulated lenses from the TDLMC previously used to validate the hierarchical framework, our analytical profiles are less cored, with slight differences in shape at larger radii. 

We now have a population of mock lens profiles which are comparable to both observed populations, except that the only difference between our SLACS-like and TDCOSMO-like populations is that we have placed the sources at two specific source redshifts. By drawing a random subset from this population, we created a set of lenses to be modeled while still maintaining confidence that our set is realistic.

\subsection{Creating mock images and fits}\label{ssec:mockimages}
We followed the setup of \citet{TDCOSMO7} to create HST-like mock images using \texttt{lenstronomy}\footnote{
    \url{https://github.com/sibirrer/lenstronomy}}
\citep{Birrer18, Birrer21a}. The images were created on a 151x151 pixel grid with a resolution of 0.08 arcseconds and were convolved by the same PSF as used in the TDLMC \citep{Ding21b}. We included noise using a zero-point of 26 mag, with an exposure corresponding to 5400s and sky brightness of 22.3 mag/arcsec$^2$ with a read out noise of 21 $e^-$, characteristic of the HST WFC3 camera in the F160W filter \citep{Dressel12}.

\begin{figure}
    \centering
    \includegraphics[width=\linewidth]{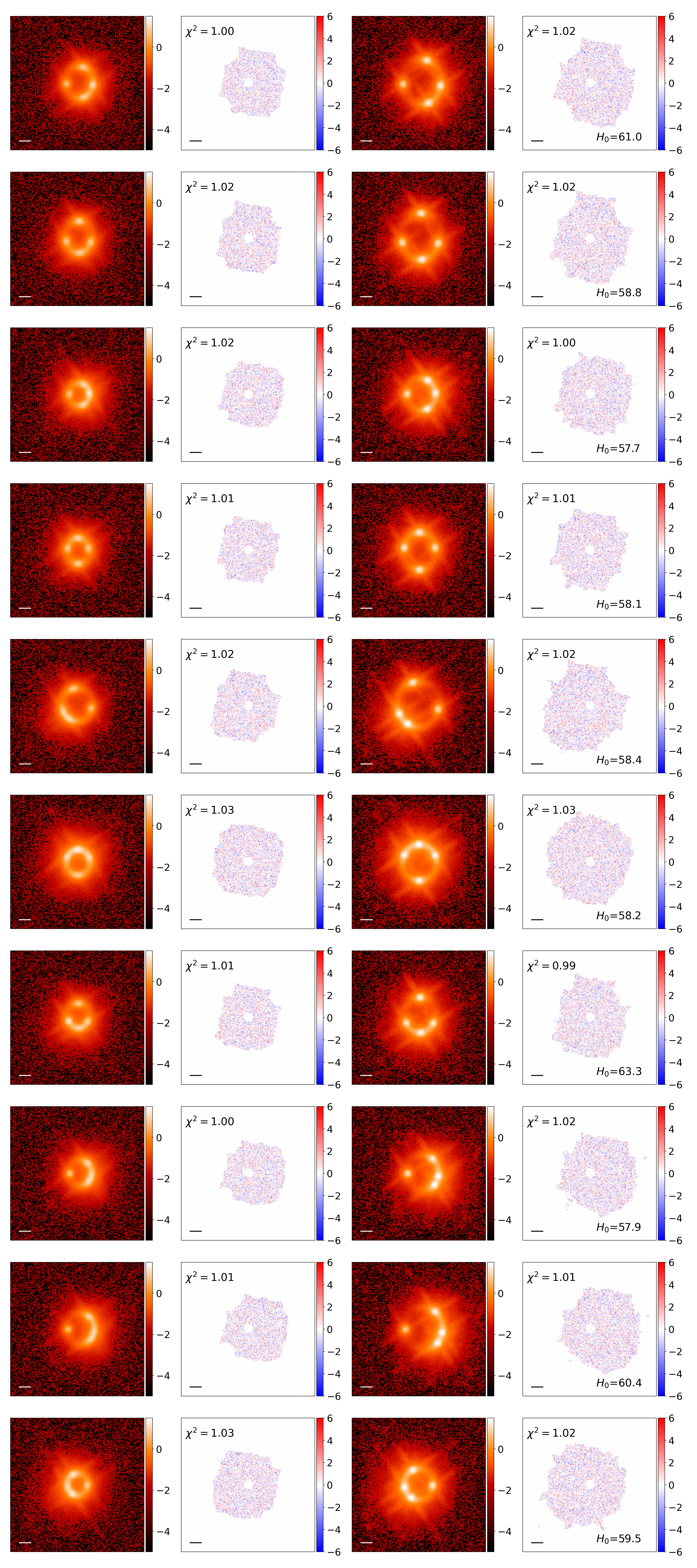}
    \caption{Images (log brightness) and residuals (in units of $\sigma$) of 10 systems at 2 source redshifts. Left: SLACS analog with $z_s=0.6$ and no point sources, Right: TDCOSMO analog with $z_s=2.0$. Lower-left scale bars correspond to 1\arcsec.}
    \label{fig:10_images}
\end{figure}

For each profile, two images were created: one SLACS-like image with $z_s=0.6$ and one TDCOSMO-like image with $z_s=2.0$. In all cases, we restricted our analysis to quadruply imaged configurations. The source was placed randomly within the caustic area in the same location for both source redshifts, which required rescaling the source position by the same factor as the increased caustic size for the higher-$z_s$ TDCOSMO-like set. The source light was a circular S\'ersic profile with $n=3$, $\theta_{\rm eff}=0.1\arcsec$ and a total unlensed magnitude of 20.5. Lacking an AGN in reality, the SLACS-like set was not given a point source or time delays. Meanwhile for the TDCOSMO-like set we included a central point source representing an AGN with an unlensed magnitude of 21. These magnitudes were selected to match the observed contrast between the quasar and host galaxy \citep{TDCOSMO7} and correspond to a system approximately 0.5 mag brighter than the mean intrinsic magnitude of the TDCOSMO lenses measured by \citep{Ding21a}. Time delays were included, centered on their true values, with an uncertainty of 2\% or 1 day  (whichever is larger) for likelihood calculation. For both sets of lenses, light from the lens galaxy was not included in an effort to focus the exploration on the changing Einstein radius rather than any possible effects of imperfect lens light subtraction. The end result for each lens was two mock high-quality HST images with different Einstein radii probing different parts of the same lens profile. Ten example lenses are shown in Fig. \ref{fig:10_images}.

The images were fit via Particle Swarm Optimization \citep[PSO, ][]{Kennedy95,Shi98} using \texttt{lenstronomy} with a S\'ersic model for the source light and a Power-law Elliptical Mass Distribution (PEMD) + external shear model for the mass. Because external shear was included in the fit model but absent in the mock creation, the fit model has more azimuthal freedom than the input. This reduces the possibility for a bias to arise due to insufficient azimuthal freedom \citep{Kochanek21}.  The central 0.4\arcsec were masked in the fitting process to stand in for the sometimes-uncertain nature of lens light subtraction in the centermost region, remaining agnostic to the possiblity of a fifth image (although our PEMD model does not produce a fifth image). We also masked all pixels with intensity below twice the background level, as in \citet{TDCOSMO7}. Because this work uses mocks where the source properties are known, we were able to initialize a preliminary fitting using the correct source, but we did not fix the source to the truth during the fitting procedure. We fixed the slope to isothermal for this preliminary fit. A secondary fit started from this result and allows all parameters to vary. From the second result, a Markov Chain Monte Carlo (MCMC) via \texttt{emcee} \citep{Goodman10,Foreman-Mackey13} was used to sample the posterior distribution of the lens parameters and estimate uncertainties.

All systems were successfully fit with the PEMD+shear models, with reduced $\chi^2<1.06$ in all cases. Before accounting for the MSD, these fits recovered $H_0$ with a median of 58.6 km s$^{-1}$ Mpc$^{-1}$, which is biased by approximately 16\% from the input value, with approximately 3\% scatter. An example comparison between the input and fit profiles for a single pair of systems is given in Appendix \ref{sec:kappa_pemdfit}. The next steps were to add kinematic information to break the MSD and combine the posteriors together hierarchically.

\subsection{Kinematics implementation}\label{ssec:kinematics}
Stellar kinematic information is used in conjunction with lensing information to break the MSD. To do the same in this work, we had to create for each lens a predicted velocity dispersion corresponding to a mock aperture measurement. 

The hierarchical framework uses spherical Jeans modeling as detailed in \citet{Binney82,TDCOSMO4}.  This approximation is commonly employed in lens modeling \citep[e.g.,][]{Treu02a,Koopmans03,Suyu10,Sonnenfeld12,Wong17,Birrer19,Rusu20} and calculates the velocity dispersion in the limit of a relaxed, spherically symmetric, nonrotating system. Radial ($\sigma_r^2$) and tangential dispersions ($\sigma_t^2$) are described by the spherical Jeans equation:
\begin{equation} \label{eq:jeans_eq}
    \frac{\partial(\rho_{\ast}(r)\sigma_r^2(r))}{\partial r} + \frac{2\beta_{\rm ani}(r)\rho_{\ast}(r)\sigma_r^2(r)}{r} = - \rho_{\ast}(r) \frac{\partial \Phi(r)}{\partial r}
\end{equation}
for a given stellar distribution $\rho_{\ast}(r)$ and gravitational potential $\Phi(r)$ where the anisotropy $\beta_{\rm ani}(r)$ is defined as
\begin{equation}
    \beta_{\rm ani}(r)\equiv1-\frac{\sigma^2_{t}(r)}{\sigma^2_r(r)}.
\end{equation} 
The luminosity-weighted projected velocity dispersion is given as
\begin{equation} \label{eq:proj_v}
    \Sigma_\ast(R)\sigma_s^2(R) = 2 \int_{R}^{\infty} \left( 1- \beta_{\rm ani}(r) \frac{R^2}{r^2}\right) \frac{\rho_{\ast}(r)\sigma_r^2(r) r {\rm d}r}{\sqrt{r^2-R^2}},
\end{equation}
where $R$ is the 2D projected radius and $\Sigma_{\ast}(R)$ is the 2D projected stellar density. Finally, to produce an observable measure, one needs to integrate this quantity over an aperture $\mathscr{A}$ and convolve by a PSF $\mathscr{P}$:
\begin{equation} \label{eq:sig_p}
    (\sigma^P)^2 = \frac{\int_{\mathscr{A}} \left[ \Sigma_{\ast}(R)\sigma_s^2(R) \ast \mathscr{P}\right]{\rm d} A}{\int_{\mathscr{A}} \left[ \Sigma_{\ast}(R) \ast \mathscr{P}\right]{\rm d} A}.
\end{equation}

All told, one can calculate the observed aperture velocity dispersion given a mass distribution, a light distribution, and a description of the orbital anisotropy. One common parameterization 
of the anisotropy is to use 
\begin{equation}
    \beta_{\rm ani}(r)=\frac{r^2}{r_{\rm ani}^2+r^2},
\end{equation}
which describes the orbits as isotropic at small radii and radial at large radii, with transition radius $r_{\rm ani}$ \citep{Osipkov79,Merritt85}. It is also useful for our purposes to describe this transition radius relative to the effective radius using the anisotropy parameter 
\begin{equation}
    a_{\rm ani} \equiv \frac{r_{\rm ani}}{r_{\rm eff}}.
\end{equation}

The spherical Jeans framework has the advantage that it is tractable to calculate quickly for a large number of models, although this is a simplification of the true dynamics of galaxies. We initially experimented with using the more general Jeans Anisotropic Modeling (JAM) \citep{Cappellari08, Cappellari20, Shajib19b}, which is starting to be explored for lens systems \citep[e.g. using mock JWST kinematics, ][]{Yildirim20,TDCOSMO8}, to calculate an input velocity dispersion while keeping the hierarchical modeling framework as spherical Jeans. However, there are subtleties regarding the differences between the two methods which make interpretation difficult. The main challenge is that the anisotropy is parameterized differently in JAM ($\beta=1-\sigma^2_{z}/\sigma^2_r$, instead with respect to the z-coordinate of a cylindrical coordinate system), which can change the aperture-measured velocity dispersion. Therefore, the experiment would not be a test of the hierarchical framework's multiple-redshift capabilities, but instead a test of the accuracy of anisotropy assumptions. Determination of an optimal description of anisotropy for lensing work proves to be a significantly more involved task than the scope of this work, and requires a focused study of its own. In the future, it would be preferable to upgrade the hierarchical framework to work with JAM, but the challenge is that JAM is too computationally expensive to sample within an MCMC. Another project is currently underway to attempt to speed up this calculation through the use of a neural network to precalculate the posterior space for MCMC sampling (paper in prep.). In this present work, we simply used spherical Jeans modeling in both the input and in the hierarchical fitting.

Regarding previous work using Jeans modeling within the hierarchical framework, we note that the kinematics calculation of \citet{TDCOSMO4,TDCOSMO5} approximated the light distribution as a Hernquist profile for the $\rho_{\ast}$ tracer population in Eq. \ref{eq:jeans_eq}. Since our light distribution is a Chameleon profile, we updated \texttt{lenstronomy} so that it is capable of calculating kinematics for Chameleon profiles (see Appendix \ref{sec:cham_troub} for details). This way, we avoid any potential bias that could come from a mismatch between the true light distribution and the one used to calculate the kinematics.

The specific kinematic setup used in this work is the same throughout. For each lens we calculated mock velocity dispersions within 3 apertures with radii of 0.67\arcsec, 1.33\arcsec, and 2\arcsec, convolved with a seeing of 0.7\arcsec. Each dispersion was given an uncertainty of $10\%$. For real systems, IFU spectroscopy is capable of measuring 2D kinematic constraints \citep[such as in MaNGA e.g., ][]{Bevacqua22} which correspond to multiple aperture constraints \citep{Shajib20}. Some of the existing lens set already has such constraints via MUSE, typically made with $\simeq5\%$ precision (see Table E.1 of TDC4 for the SLACS set). Furthermore, resolved kinematics should be possible with JWST \citep{Yildirim20} and adaptive-optics IFU data such as MUSE \citep{Bacon10}, KCWI \citep{Morrissey18} and OSIRIS \citep{Larkin06}, providing even higher precision constraints for some subset of the lens sample. Indeed, IFU kinematics are anticipated within the TDCOSMO forecast using the hierarchical framework \citep{TDCOSMO5}. As such, our kinematic setup uses a conservative estimate of the constraining power of real observations. For our mock kinematics, we must set an input anisotropy $a_{\rm ani}$. For real systems $a_{\rm ani}$ must take on a range of values but the exact physical distribution is not well-known. To focus the experiment on the effect of changing impact parameter rather than the parameterization of anisotropy, we set input $a_{\rm ani}=1$ for all systems. The limitations of this assumption are further discussed in Sect. \ref{ssec:mock_limits}. This describes the mock measured velocity dispersions calculated for the true profile.

We must also calculate the predicted velocity dispersions from the lensing model. This was done using the same spherical Jeans modeling for the PEMD fit profile, resulting in a measured velocity dispersion for each aperture. The posteriors of the lensing fit were sampled, with uncertainties in the fit slope and Einstein radius included. We also introduced a $10\%$ uncertainty to the effective radius in this sampling. Considering the effect of the MST on velocity dispersion, we note that the infinite uniform sheet term in Eq. \ref{eq:mst} does not affect kinematics. As such, the effect of the MST on the velocity dispersion is a modification by a constant factor of $\sqrt{\lambda_{\rm int}}$ which follows from the application of the coefficient $\lambda_{\rm int}$ to the density in Eq. \ref{eq:proj_v} (or from the simple scaling that $v^2\propto GM/R$). The hierarchical framework seeks to match the predicted and measured velocity dispersions, sampling over the parameters which describe the population-level $a_{\rm ani}$ and $\lambda_{\rm int}$. 

A summary table of the parameters used for mock lens creation is made available. See Appendix \ref{sec:bounds}.

\section{Results} \label{sec:hierarc_results}
We now present results for an ensemble of lens systems. 
We used the publicly available \texttt{hierArc} package\footnote{
    \url{https://github.com/sibirrer/hierArc}} \citep{TDCOSMO4}
to sample over the posteriors from the PEMD fits to each system. From the lensing posteriors and kinematic information, \texttt{hierArc} samples for $H_0$, using a linear parameterization of the population-scale MST (Eq. \ref{eq:mst_parameterization}) with intercept $\lambda_{\rm int,0}$ and slope $\alpha_{\lambda}$, with a spread of $\sigma(\lambda_{\rm int})$. Additionally, the distribution of anisotropy is described as Gaussian with median $\langle a_{\rm ani}\rangle$ and standard deviation $\sigma(a_{\rm ani})$. This work uses uniform priors for all of these parameters with bounds listed in Appendix \ref{sec:bounds}. One can compare the results from including just the lenses from the one or the other source redshift, or a combination using the populations from both source redshifts.

\subsection{20 systems} \label{ssec:20systems}
\begin{figure*}
    \centering
    \includegraphics[width=0.95\textwidth]{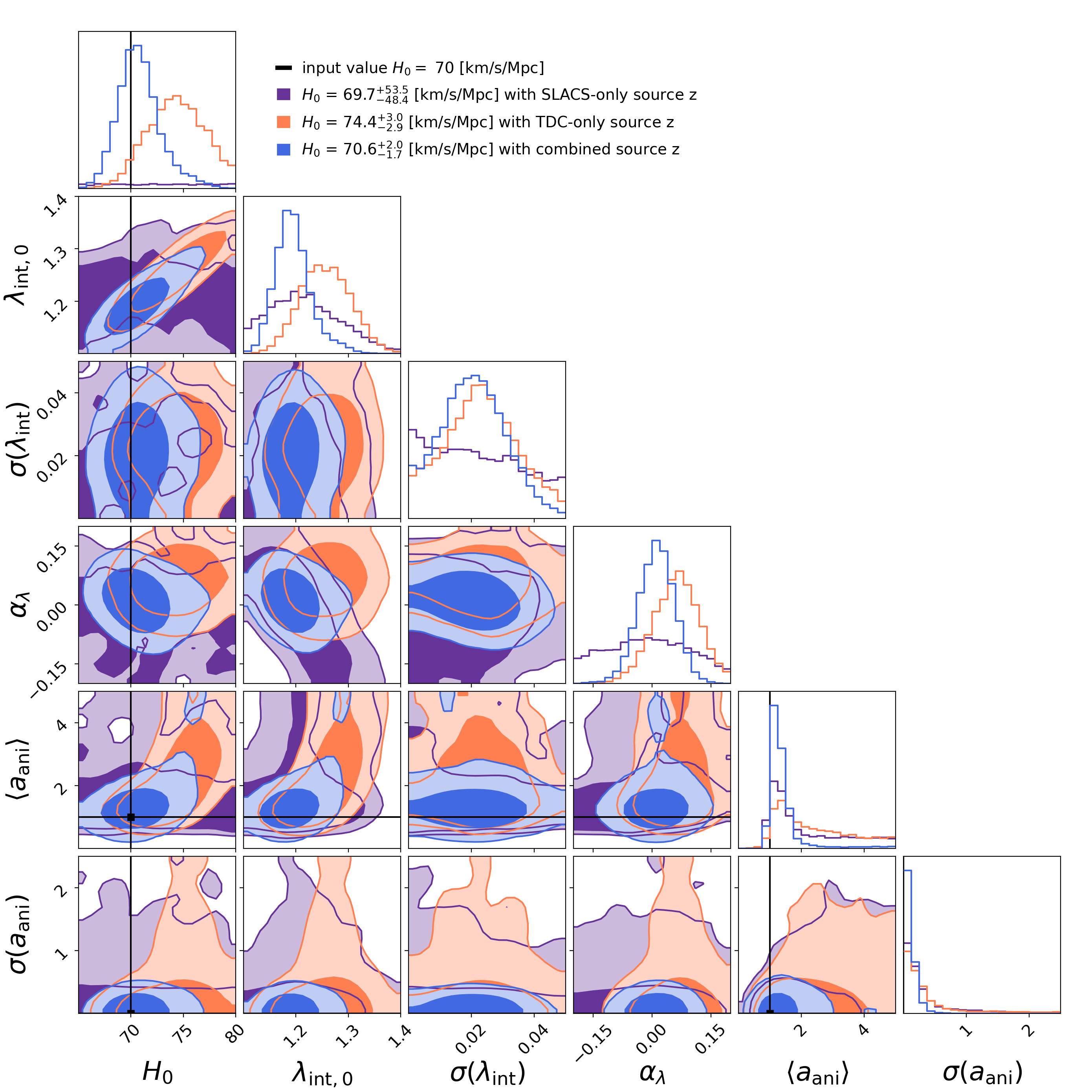}
    \caption{Results of the hierarchical framework using 20 Chameleon+NFW lenses for the population with SLACS source redshift and no time delays (purple), the population with TDCOSMO source redshift (orange) and the combination of both populations (blue).}
    \label{fig:hierarc20pop}
\end{figure*}

The results for 20 Chameleon+NFW lenses are plotted in Fig. \ref{fig:hierarc20pop}. A comparison of the SLACS-only population, the TDCOSMO-only population, and the combined populations indicates that the combination can increase the precision and accuracy on the final value of $H_0$, even though the systems span different ranges of $\theta_{\rm eff}/\theta_{\rm Ein}$. 

The TDCOSMO-only set recovered a median $H_0$ which is higher than the fiducial value of $H_0$ by approximately $6.2\%$ ($1.5\sigma$). The broad width of the posterior is largely due to the degeneracy with $\langle a_{\rm ani} \rangle$, which was poorly constrained from a single-redshift set of lenses.

The SLACS-only population had no point sources, therefore no time delays; as a result, it could not place a constraint on $H_0$ alone, as evidenced by the wide and unconstrained purple contours in the first column of Fig. \ref{fig:hierarc20pop}. Rather, the contribution of these systems is to inform the parameters other than $H_0$: those concerning $\lambda_{\rm int}$ and $a_{\rm ani}$. This information is combined with the TDCOSMO systems (which themselves bring information about the $H_0$ posterior) to break the MSD. This combination enabled the fit to recover an unbiased $H_0$ value of $70.6^{+2.0}_{-1.7}$ km s$^{-1}$ Mpc$^{-1}$. As a particularly nice visual example, the panel which shows $\alpha_{\lambda}$ and $\lambda_{\rm int,0}$ (middle row, second column) illustrates how these parameters, while degenerate within the SLACS posteriors, were nonetheless able to inform the TDCOSMO set to recover an accurate description of $\lambda_{\rm int,0}$. As in TDC4, the combined-population result was shifted downward, but still consistent with the TDCOSMO-only value of $H_0$.

\subsection{40 systems}\label{ssec:40systems}
While 20 systems is a reasonable short-term goal of TDCOSMO, a larger number of lenses will soon become available with the advent of LSST, Euclid, and larger telescopes. Increased sample size should increase the precision on $H_0$, but intrinsic scatter in $\lambda_{\rm int}$ may inhibit this precision increase. We took this opportunity to test if an increased number of lens systems within the hierarchical framework can correctly quantify and account for the intrinsic scatter in $\lambda_{\rm int}$. As such, we doubled the number of systems to 40 at each redshift.

\begin{figure*}
    \centering
    \includegraphics[width=0.95\textwidth]{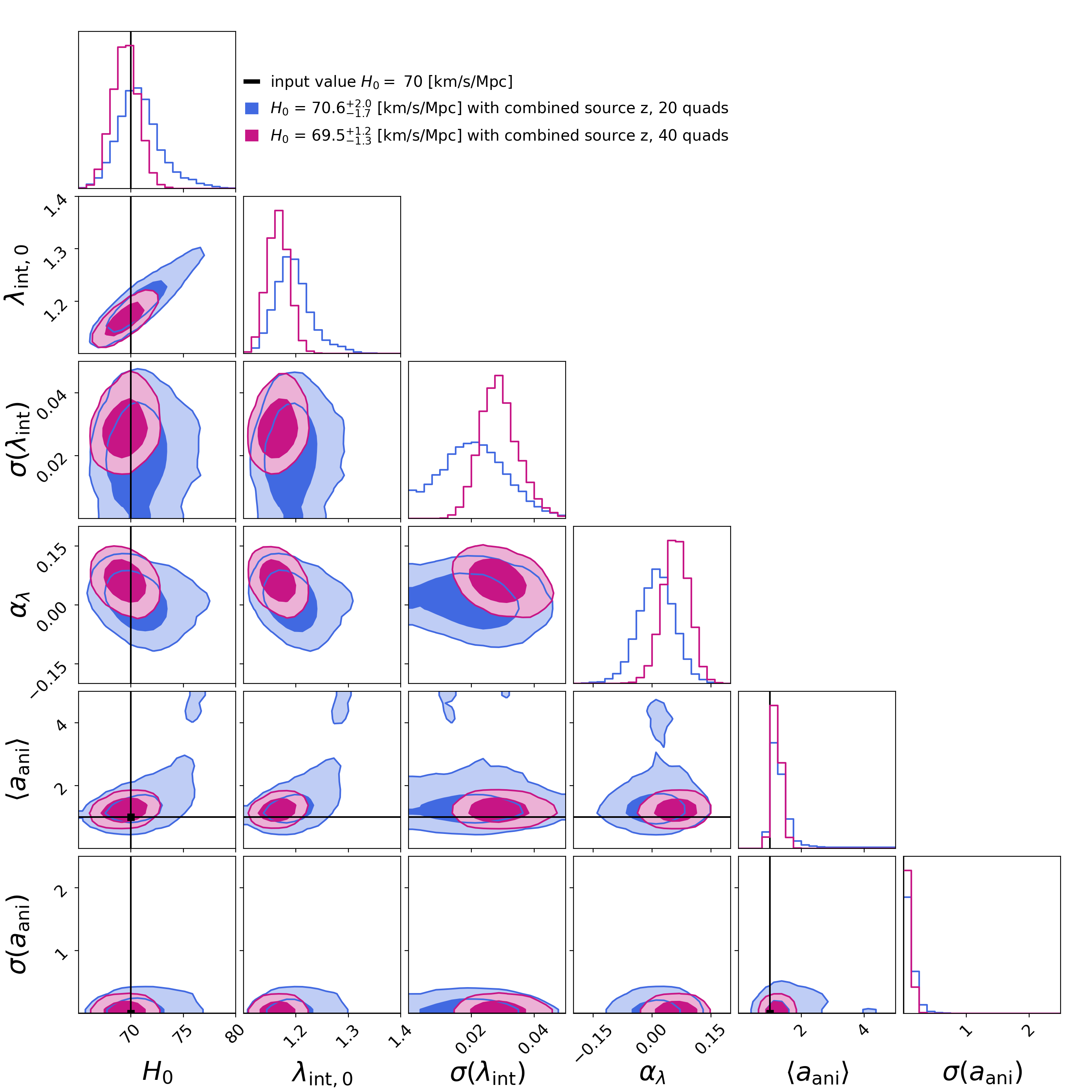}
    \caption{Effect of the number of lenses: the hierarchical result of a population of Chameleon+NFW lenses, combining the fits from the TDCOSMO and SLACS source redshifts. The blue is the same as that in Fig. \ref{fig:hierarc20pop}, using 20 lenses from each redshift. The pink distribution corresponds to the same setup, but with 40 lenses each.}
    \label{fig:hierarc40pop}
\end{figure*}

With 40 systems at each redshift, we recovered the results in Fig. \ref{fig:hierarc40pop}. As intended, the combined SLACS+TDCOSMO value errors did indeed shrink, zeroing in on the correct value of $H_0$, with a recovered value of $69.5^{+1.2}_{-1.3}$ km s$^{-1}$ Mpc$^{-1}$.

\subsection{Single-redshift test} \label{ssec:single_z}
We were curious about the importance of probing a range of redshifts, that is to say spread over a greater range of impact parameter. To discuss this, we must compare against the case in which the lenses come from the same redshift. In Appendix \ref{sec:PSSLACS}, we show the results of a test where we set the systems to the same redshift and ran \texttt{hierArc}. With 20 time delay and 20 non-time-delay lenses all at the SLACS $z_s$, the result was $H_0=69.7^{+1.6}_{-1.6}$ km s$^{-1}$ Mpc$^{-1}$.

\section{Discussion}\label{sec:discussion}
The goal of this project has been to test the systematics of the hierarchical framework. The framework has already been tested on a single sample of time-delay lens systems from hydrodynamical simulations; we have expanded this analysis to a different input model using two-component analytical profiles consistent with the observed lens population. Furthermore, we have expanded the analysis to test the effect of combining the sample with an analogous population of non-time-delay lenses with different source redshifts. 

\subsection{Prehierarchical result}
\begin{figure}
    \centering
    \includegraphics[width=0.95\linewidth]{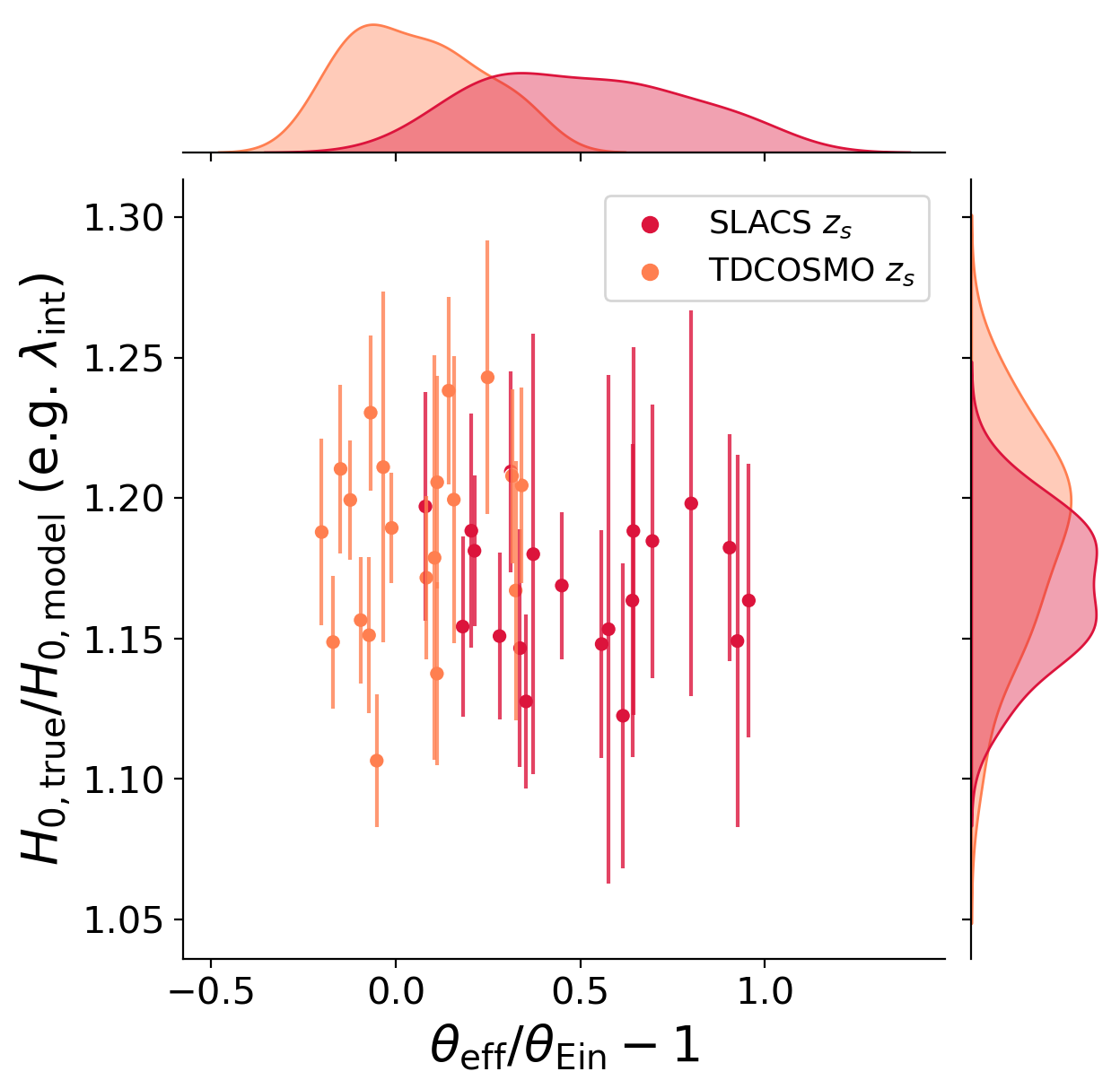}
    \caption{Lensing-only values of $H_0$ (i.e., $\lambda_{\rm int}$) as a function of radial impact parameter defined in Eq. \ref{eq:mst_parameterization}, which assumes this relationship is described as linear with Gaussian scatter. The SLACS-$z_s$ systems are indicated in red rather than purple because they are alternate versions of the same systems but with point sources and time delays (such that a value for $H_0$ can be recovered and plotted).
    }
    \label{fig:lambda_scaling}
\end{figure}

To better understand the results of the hierarchical fitting, it can be useful to compare to the lensing results alone, before the MST is accounted for. Figure \ref{fig:lambda_scaling} shows the lensing-only recovered value of $H_0$ for 20 systems at each $z_s$, expressed as $H_{0, \rm true}/H_{0, \rm model}=\lambda_{\rm int}$, as a function of the radial scaling parameterization $\theta_{\rm eff}/\theta_{\rm Ein}-1$. We note that since the SLACS-like systems have no point sources, they do not recover $H_0$ values, and so we have instead plotted the results from the test where the same 20 lenses were given point sources in Sect. \ref{ssec:single_z}. The hierarchical parameters of $\lambda_{\rm int,0}$, $\sigma(\lambda_{\rm int})$, and $\alpha_\lambda$ reflect the summary statistics of a linear-fit description of this plot. We can see $\lambda_{\rm int}$ has a median of approximately 1.18 and scatter of 0.03, which should be approximately the truth values of $\lambda_{\rm int,0}$ and $\sigma(\lambda_{\rm int})$, assuming $\alpha_\lambda=0$ (which seems reasonable given the weak correlation, Pearson $R=-0.18$).

The lensing data is supplemented with kinematic data, and so it can be enlightening to see the velocity dispersion values from the model fits to the lensing data. Before any MST correction was applied, and setting $a_{\rm ani}$ to the truth value of 1, 
we found that all fits from both systems recovered model velocity dispersions ($\sigma^P$) which were approximately $7.5\%$ lower than the true values. 
For individual systems this bias value is contained within the uncertainties of individual $\sigma^P$ measurements ($10\%$ uncertainties), but at a population level systems require an average $\lambda_{\rm int}>1$ to reproduce the (observed) lens velocity dispersion compared to the power-law model. This illustrates the necessity to use information from the population of lenses rather than individually.

\subsection{Performance of the hierarchical framework}

We first consider the capacity of the hierarchical framework to reproduce $H_0$ from a population at a single source redshift and therefore a narrow range of $\theta_{\rm eff}/\theta_{\rm Ein}$. Considering the case with 20 systems using only the TDCOSMO-like source redshift (orange set in Fig. \ref{fig:hierarc20pop}), the spread in anisotropy radius $\langle a_{\rm ani}\rangle$ propagates into $4.4\%$ uncertainty in $H_0$, with a true value which lies within the 95\% confidence interval of of the posterior. When expanded to 40 systems (not plotted), this resulted in $H_0=72.3^{+2.3}_{-1.7}$ km s$^{-1}$ Mpc$^{-1}$, again with the true value within the 95\% confidence interval, but now with an uncertainty of only $2.8\%$. It is interesting that the result lies approximately $1.5\sigma$ away from the truth in both cases, even as the value of $\sigma$ narrows. This may indicate that discrepancies still could exist within the assumptions in the framework at the $\simeq1\sigma$ level, but $1.5\sigma$ is not severe enough to claim significance.  Nevertheless, this precision represents a substantial improvement over the MST-biased results from lens modeling alone, which had a median bias of $16\%$ below the fiducial value. This illustrates the success of the hierarchical framework in breaking the MSD for a single population.

We further consider the capacity of the hierarchical framework to account for a change in the impact parameter for a set of otherwise similar lenses. The SLACS-like population is identical to the TDCOSMO-like except that they have no point sources and have a smaller $z_s$ resulting in a larger $\theta_{\rm eff}/\theta_{\rm Ein}$. Here we compare the result from the single-redshift population to that of the combined population. The combined-population constraints on $\langle a_{\rm ani}\rangle$ are more informative than either individual set. The degeneracy between $\langle a_{\rm ani}\rangle$ and $H_0$, visible in the TDCOSMO-only set of Fig. \ref{fig:hierarc20pop}, shows how this improved anisotropy constraint leads to a more precise $H_0$ ($2.6\%$ scatter for 20 systems).  The accuracy is also improved: our combined set for 20 systems was centered just $0.3\sigma$ away from the correct value. When we increased the number of systems to 40, the precision and accuracy improved further, to a final value of $69.5^{+1.2}_{-1.3}$ km s$^{-1}$ Mpc$^{-1}$ ($0.7\%$ median offset, $1.8\%$ scatter). The increase in accuracy comes primarily from tighter constraints on $\lambda_{\rm int,0}$ as the large number of systems is better able to probe the systematic behavior. Nonzero $\sigma(\lambda_{\rm int})$ was also recovered, meaning the framework was able to capture the presence of the intrinsic variance in the input population. Based on Fig. \ref{fig:lambda_scaling}, this scatter is anticipated to be approximately 3\%, which was well-matched by the recovered value of $\sigma(\lambda_{\rm int})$. Because the hierarchical framework explicitly captures this scatter through $\sigma(\lambda_{\rm int})$, intrinsic scatter in the distribution of $\lambda_{\rm int}$ does not represent a systematic floor for $H_0$, that is to say it does not preclude the recovery of $H_0$ at a better precision than the intrinsic scatter, and so one can expect the precision to improve as $\sqrt(N)$. Importantly, the correct distribution of $a_{\rm ani}$ was recovered, which forced $H_0$ to the correct value. The power to constrain $a_{\rm ani}$ comes from having multiple-aperture kinematic measurements, providing a probe of the behavior of velocity dispersion with radius. In summary, we confirm that the hierarchical framework is capable of incorporating lens populations from different source redshifts.

When considering the test where all lenses are placed at the same redshift, we find that the framework successfully recovered $H_0$ in this test as well. We conclude that the method can result in an accurate value of $H_0$ when using either a set of lenses at the same redshift or combining sets from different redshifts.

\citet{TDCOSMO5} forecasted the improvement in precision achieved within the hierarchical framework by increasing the size of the lens population, assuming improved constraining power of the velocity dispersion from integral field unit (IFU) spectroscopic data. Unlike this work, \citet{TDCOSMO5} did not create and fit mocks, but simply assumed Gaussian constraining power for each individual lens. With such a setup, they found that with 40 time-delay lenses and up to 200 non-time-delay lenses, the precision can be expected to increase to $1.2-1.5\%$, depending on the exact setup. Our result using simulated mocks from start to finish found $1.8\%$ precision with 40 time-delay and 40 non-time-delay lenses, consistent with this forecast.

We note that the general behavior of having a lower value of $H_0$ for the combined sample than the TDCOSMO-only sample qualitatively resembles the result from TDC4, where the same hierarchical framework was used on the observed lenses. TDC4 found $H_0=74.5^{+5.6}_{-6.1}$ km s$^{-1}$ Mpc$^{-1}$ when using 7 TDCOSMO lenses alone, but found that the value shifted to $67.4^{+4.1}_{-3.2}$ km s$^{-1}$ Mpc$^{-1}$ when incorporating 33 SLACS lenses. Speaking in terms of the downward uncertainty of the TDCOSMO-only set, the value shifted by $1.1\sigma$, whereas for this work we saw a similar $1.3\sigma$ shift. While this work cannot confirm or refute the mutual similarity between the real lens populations, it is reassuring to see consistent results with the real population, as it affirms that the observed shift is not necessarily an indicator of dissimilar populations. Such mutual similarity between the populations is the key assumption required to make use of the SLACS lenses hierarchically; this work finds no evidence against this assumption.

\subsection{Limitations}
\subsubsection{Mock data}\label{ssec:mock_limits}

The results of this project look optimistic for the hierarchical framework, but we would be remiss to not discuss the differences between this work and TDC4, as well as some of the present limitations of our method. Firstly, some choices made for the mock data quality may differ slightly from the real lens population. This work has a larger number of systems compared to TDC4, which at the time included 7 TDCOSMO lenses and 33 SLACS lenses of which only 9 had multiple-aperture velocity dispersion measurements, with approximately $5\%$ precision on $\sigma^P$. Here we used 20 (or 40) of each set, all of which have 3 kinematic apertures with $10\%$ precision on $\sigma^P$. Image data quality in this work (e.g., source brightness, noise, exposure time, etc.) was modeled after the high S/N template from \citet{TDCOSMO7}, and as such are higher-than-average quality compared with current single-band data. We also used transparent lenses, which assumes perfect lens light subtraction. On the other hand, we used single-band images while the observed multiband imaging add additional constraining power and can help deblend lens and source light \citep[e.g.,][]{Shajib20,Schmidt22}. This work is intended to be forward-looking, and the sample of real lenses will grow over time to better match these settings in the future. 

By design, this work uses only two source redshifts and one lens redshift to more accurately probe what is thought to be the key difference between the SLACS and TDCOSMO lenses. The real lens population spans a range of $z_s$ and $z_d$, but since our populations already share similar distributions of $\theta_{\rm eff}/\theta_{\rm Ein}$, implementing this change would be unlikely to make a significant difference. However, if there are different selection effects between the observed SLACS and TDCOSMO samples, or changes to the global profile shape associated with galaxy evolution between the two sample lens redshifts, these effects are not captured in this work.

Regarding our choice of profile, the Chameleon+NFW profiles match many aspects of the observed systems, but may not perfectly describe the truth. Our SLACS-like $\xi$ distribution has less scatter and is more isothermal than the observed set. Additionally, the fact that our PL lensing fits resulted in a different $H_0$ value than the truth may indicate a different behavior than the real lens population, which find similar values of $H_0$ whether a PL or composite model is used \citep{TDCOSMO1}. However, to call this a discrepancy assumes that fitting our systems with composite models would return the correct $H_0$ (no MST), which we do not necessarily claim. At any rate, we can state that the hierarchical framework successfully accounts for the MSD using either our analytical mocks or the hydro-simulated mocks of the TDLMC, supporting the claim that it is reliable for the true population of (unknown) lens profiles.

A salient difference between our mocks and real galaxies is their elliptical shape, while real galaxies harbor more complex azimuthal structures. The level of impact these structures can have on $H_0$ inference is debated.  Early-type galaxies can be triaxial \citep{vandeVen09, Chang13, Weijmans14}, and can have 2D mass projections which can change in ellipticity or position angle with radius \citep{Kormendy09}. We initially sought to create lenses from 3D triaxial profiles for this work, but challenges with the kinematics calculation forced us to simplify to 2D elliptical profiles. Furthermore, galaxies have been known to have "disky" or "boxy" isophotes, corresponding to an $n=4$ multipole perturbation to their mass distributions \citep{Hao06}. Finally, in the $\Lambda$CDM paradigm there are expected to be a large number of subhalos within any given galactic halo which add substructure to the mass distribution \citep{TDCOSMO3}. Additional angular complexity of a lens can lead to a bias in $H_0$ when a model is not adequately azimuthally complex \citep{Kochanek21}. The effects of some of these angular structures in the context of lensing measurements of $H_0$ are explored in other works (e.g., substructure by \citet{TDCOSMO3}, multipole by \citet{TDCOSMO7}, isodensity twists by \citet{vandeVyvere22}). However, it is clear that any substructures are a subdominant factor to the elliptical shape of the mass distribution as evidenced by the success of cosmography-grade parametric lens models. As such, our simple elliptical lenses are likely a reasonable approximation for this work, for which the purpose is to probe the effect of a changing radial structure.

Along the same lines, our spherical Jeans framework used to calculate both the input and the fit velocity dispersion is a simplification. There is an inconsistency between using an elliptical mass distribution for the lens and a spherical mass distribution for the velocity. \citet{Yildirim20} found, by using JAM in the spherical limit, that this approximation did not bias the resulting $D_{\Delta t}$ measurement for their single system, although 
a formal inclusion of $\lambda_{\rm int}$ is not incorporated. This spherical assumption is common in joint lens+kinematics modeling, in which kinematics are meant only to provide a single integrated measure of mass, but may become more important as spatially resolved kinematics become more widely attainable. Perhaps of greater concern is the description of anisotropy through the single value of $a_{\rm ani}$. This parameterization is known to be a simplification which does not accurately match N-body simulations \citep{Mamon05}. Even so, it is unrealistic to expect each galaxy to have the same value of $a_{\rm ani}$ as has been set in this work as well as in \citet{TDCOSMO5}. Preliminary tests with changing the input distribution of $a_{\rm ani}$ have found that $H_0$ is still recovered well when a representative prior is used, but it is possible for biases to arise if the prior is misinformed \citep[similar to the findings of][]{Birrer16,TDCOSMO4}. There is also room to alter and explore the kinematics settings (e.g., number of apertures, radii, seeing, uncertainties) to learn if there is an optimal strategy for an ideal kinematics setup. A more thorough quantification of these effects is a target for a future paper. A future goal would be to incorporate the JAM framework into \texttt{hierArc}, which at present may be too computationally demanding. 


\subsubsection{MST parameterization}
Here we describe an ancillary result with possible implications for the theoretical description of the MST. We sought to supplement the empirical success of the linear parameterization of $\lambda_{\rm int}$ with impact parameter with an analytical description. The intention was to plot a "truth" theoretical relation in Fig. \ref{fig:lambda_scaling} representing a mapping from Chameleon+NFW to PEMD profiles, and determine if a linear parameterization was appropriate. We performed some tests using the $\xi$ values of each profile to predict the $\lambda_{\rm int}$ one would recover from a PL fit (since the PL fit should recover the same $\xi$ as the true profile). However, we found that the predicted $H_0$ using $\xi$ differs from the actual value recovered from the mock fitting by approximately 3\% (median). After considerable exploration, we find that approximately 60\% of this discrepancy can be explained by effects of ellipticity: $\xi$ is defined using a spherical lens for which the image radius is unambiguous. However, the remaining discrepancy is still not fully explained, and may be related to other degeneracies in the parameters, possibly dependent on image configuration. This discussion will be the subject of a separate publication \citep[Gomer et al. 2022 in prep][]{}. For this current work, this simply means that we cannot reliably use this $\xi$ description to represent a theoretical target. 

Finally, we mention in passing that the way the MST is parameterized physically in 3D kinematics is open to interpretation. In this work, the velocity dispersion is modified by a constant factor of $\sqrt{\lambda_{\rm int}}$, noting that the infinite sheet term in Eq. \ref{eq:mst} does not affect kinematics. An alternative description by \citep{Blum20} describes the MST as a cored profile, which is well-behaved at infinity and would leave a detectable imprint on kinematics \citep{TDCOSMO8}. Other descriptions are possible, as the deprojection of the MST is not unique. This work is confined to the first description, but future versions of \texttt{hierArc} could implement other parameterizations.

\section{Conclusion}
In this work, we conducted an experiment by creating a population of pairs of mocks of the same lens profile using 2 different source redshifts, representing the difference between the SLACS and TDCOSMO lens populations, controlling for any other possible differences. These images were constructed with two-component profiles, but fit as a power law, resulting in a $\sim15\%$ biased recovery of $H_0$. The posteriors from these fits were then combined together and supplemented with spherical Jeans velocity dispersion measurements using the hierarchical framework of \citet{TDCOSMO4}. 
Our resulting value of $H_0$ using a single population of TDCOSMO-like lenses was consistent with the fiducial $H_0$ (using 20 lenses: $4.2\%$ precision, $1.5\sigma$ median offset). When information from the SLACS-like population of lenses was used to supplement the recovery, $H_0$ was recovered more precisely and more accurately (using 20 lenses from each set: $2.6\%$ precision, $0.3\sigma$ median offset). This precision can be further improved as the number of systems grows (using 40 lenses from each set: $1.8\%$ precision, $0.4\sigma$ median offset). 

Even though the intrinsic profile was more complex than the power-law model used to fit it, we confirm that the hierarchical framework was able to combine systems from different redshifts and adjust the  parameterization of $\lambda_{\rm int}$ as a function of $\theta_{\rm eff}/\theta_{\rm Ein}$ in a way which results in an unbiased $H_0$. While there may be other possible parameterizations, the linear function with respect to scaled effective radius seems acceptable, provided the data quality is good and the number of lenses is large. 

There are still a number of open questions for the hierarchical framework, such as the effect of the distribution of $a_{\rm ani}$ and the proper way to parameterize the anisotropy of galaxies, or the effect of different 3D physical interpretations of the MSD. These questions will be further explored in future work, but this work validates the utility of the method and serves as an important first step.

\begin{acknowledgement}
The authors thank Sherry Suyu, Ak{\i}n Y{\i}ld{\i}r{\i}m, Martin Millon, Alessandro Sonnenfeld, and Matt Auger for their constructive discussion regarding this paper.

This work uses the following Python packages: Python \citep{Python1,Python2}, Astropy \citep{astropy:2013,astropy:2018}, Numpy \citep{Numpy}, Scipy \citep{scipy}, Matplotlib \citep{Matplotlib}, Pandas \citep{pandas1,pandas2}, and Seaborn \citep{seaborn}.

This project has received funding from the European Research Council (ERC) under the European Union’s Horizon 2020 research and innovation programme (grant agreement No 787886). 


\end{acknowledgement}

\bibliographystyle{aa} 
\bibliography{biblio}

\appendix

\section{Troubleshooting Chameleon profiles}\label{sec:cham_troub}

\subsection{S\'ersic-matching modification}
Initially we intended to use a S\'ersic profile directly for the baryon mass component, matching the observed family of light distributions. Unfortunately, the S\'ersic profile is computationally prohibitive to implement as a lens. Instead, we followed the practice of \citet{Dutton11} and \citet{Suyu14} to create a Chameleon profile which emulates the S\'ersic profile by using a combination of two cored isothermal profiles. In the notation of \citet{Dutton14}, Eq. \ref{eq:cham_prof} reads:
\begin{equation}
    \Sigma(R)=\frac{\Sigma_0}{1-\alpha}\left(\frac{R_0}{\sqrt{R^2+R_0^2}}-\frac{R_0}{\sqrt{R^2+(R_0/\alpha)^2}}\right).
\end{equation}

\begin{figure}
\centering
\includegraphics[width=\linewidth]{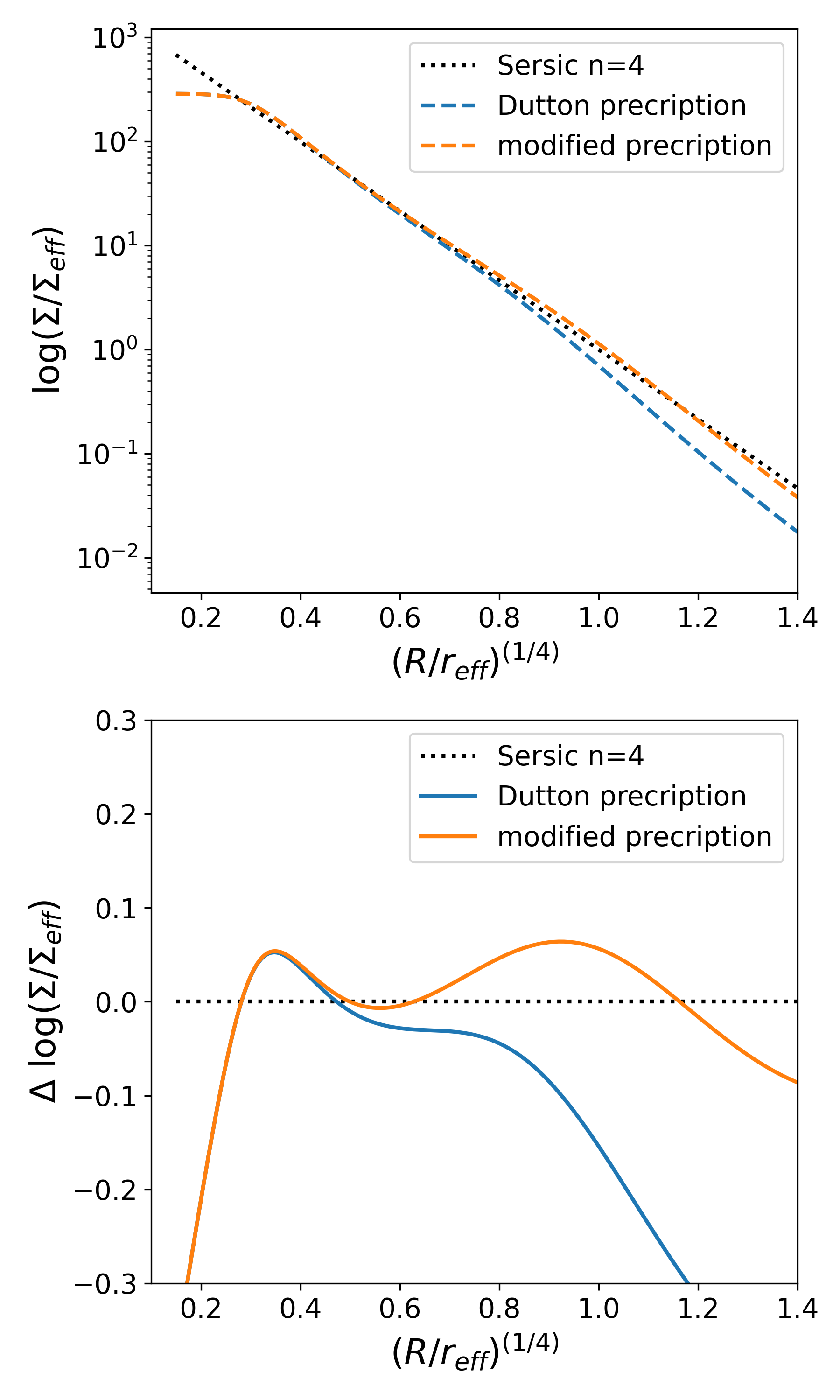}
\caption{Comparing the matching Chameleon profile to the input S\'ersic profile. The modification we have made to the prescription better reproduces the corresponding figure from \citet{Dutton11}.}
\label{fig:chamsersic}
\end{figure}

\citet{Dutton11} provide a prescription to convert a given S\'ersic profile into a Chameleon profile based on a polynomial fit. When we followed this prescription, we found we were unable to match our input S\'ersic profile to the provided precision. Instead, we found that, at least for a S\'ersic with $n=4$, we had to manually modify the ratio of the two core radii, dividing $\alpha$ by a factor of $1.5$. Only after this modification do our profiles match Fig. A2 of \citet{Dutton11}. This effect is shown in Fig. \ref{fig:chamsersic}. We have not tested the general case. We suspect that there is a transcription error in the polynomial provided within the manuscript, but it is also possible that there is an error in our implementation, so we have made available a Python notebook which shows our calculations.\footnote{\url{https://github.com/mattgomer/lenstronomy-profile-troubleshooting}} At any rate, the Chameleon profiles we use in this work are constructed from our modified prescription and match S\'ersic profiles to the precision shown in Fig. \ref{fig:chamsersic}.

\subsection{Multiple Gaussian expansion}\label{sec:mge_troub}

\begin{figure}
\centering
\includegraphics[width=\linewidth]{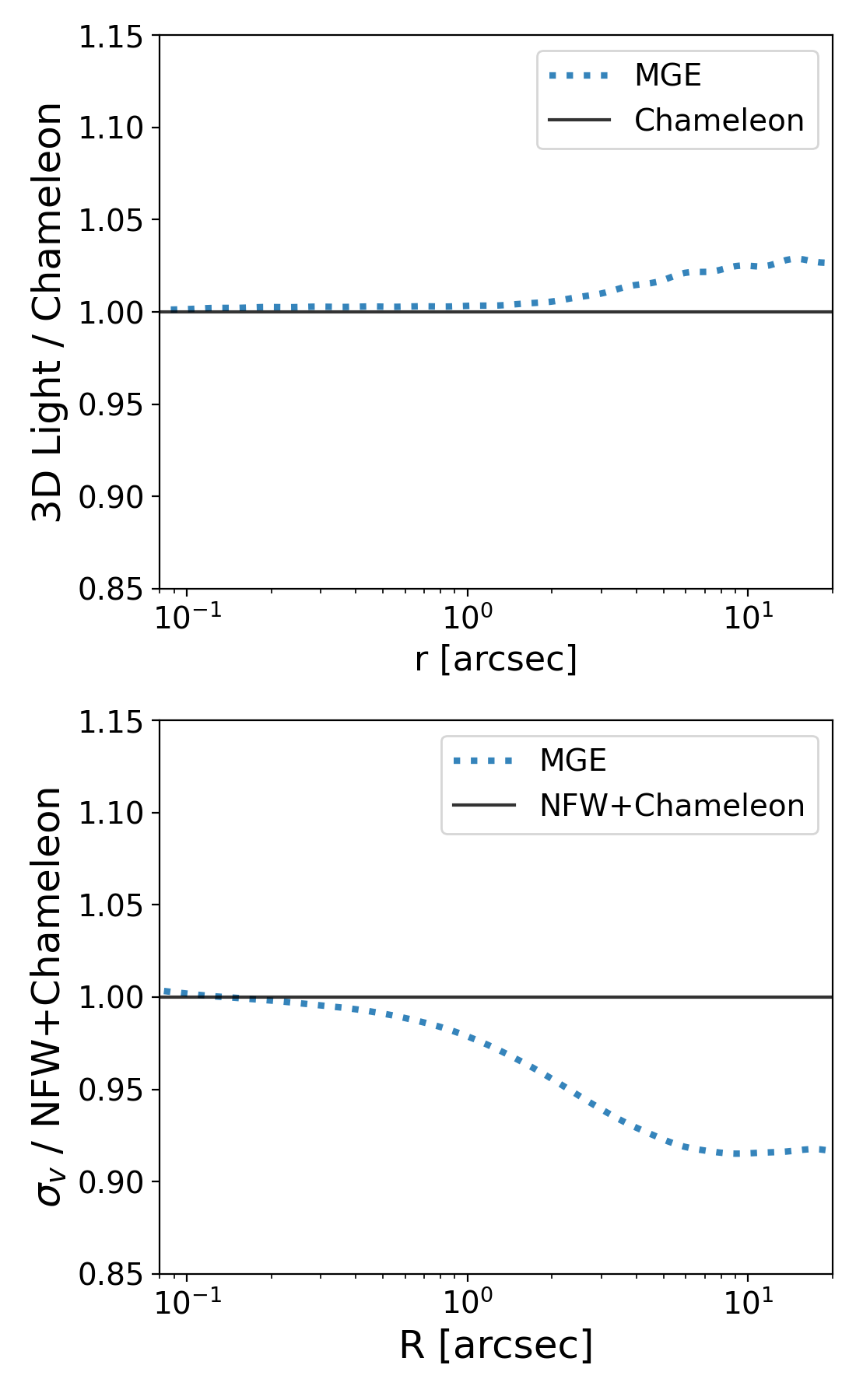}
\caption{Use of the 1D MGE introduces an excess of light in 3D at large radius for an elliptical lens. This causes the $\sigma_v$ calculation to be biased by approximately 3\% near the aperture radius of 1 arcsecond.}
\label{fig:cham_mge}
\end{figure}

While troubleshooting the kinematics calculation within lenstronomy, we discovered a discrepancy can be introduced when the light profile is modeled as a Multiple Gaussian Expansion (MGE) to approximate the profile rather than the analytical profile. The MGE is used because it is simple to deproject, and a 3D mass distribution is required for kinematics. However, \texttt{lenstronomy} uses a 1D MGE for its kinematic calculation because it assumes spherical Jeans, which can lead to inaccuracies in deprojection when starting from an elliptical mass distribution. Namely, we found that the MGE introduces an excess of light at large radii in 3D when the lens is not circular. This can introduce a bias in the calculated $\sigma_v$ at large radii, which can bias the aperture-weighed $v_{\rm ap}$ by approximately $3\%$ (see Fig. \ref{fig:cham_mge}). The effect worsens for larger radii and for more elliptical lenses (this example used $q=0.8$). To avoid this, \texttt{lenstronomy} has been updated to include the 3D profile for the Chameleon profile, such that the MGE is no longer necessary, but this serves as a cautionary tale to beware of subtle effects when calculating kinematics of lens models.

\section{Single fit example} \label{sec:kappa_pemdfit}
\begin{figure}
    \centering
    \includegraphics[width=0.95\linewidth]{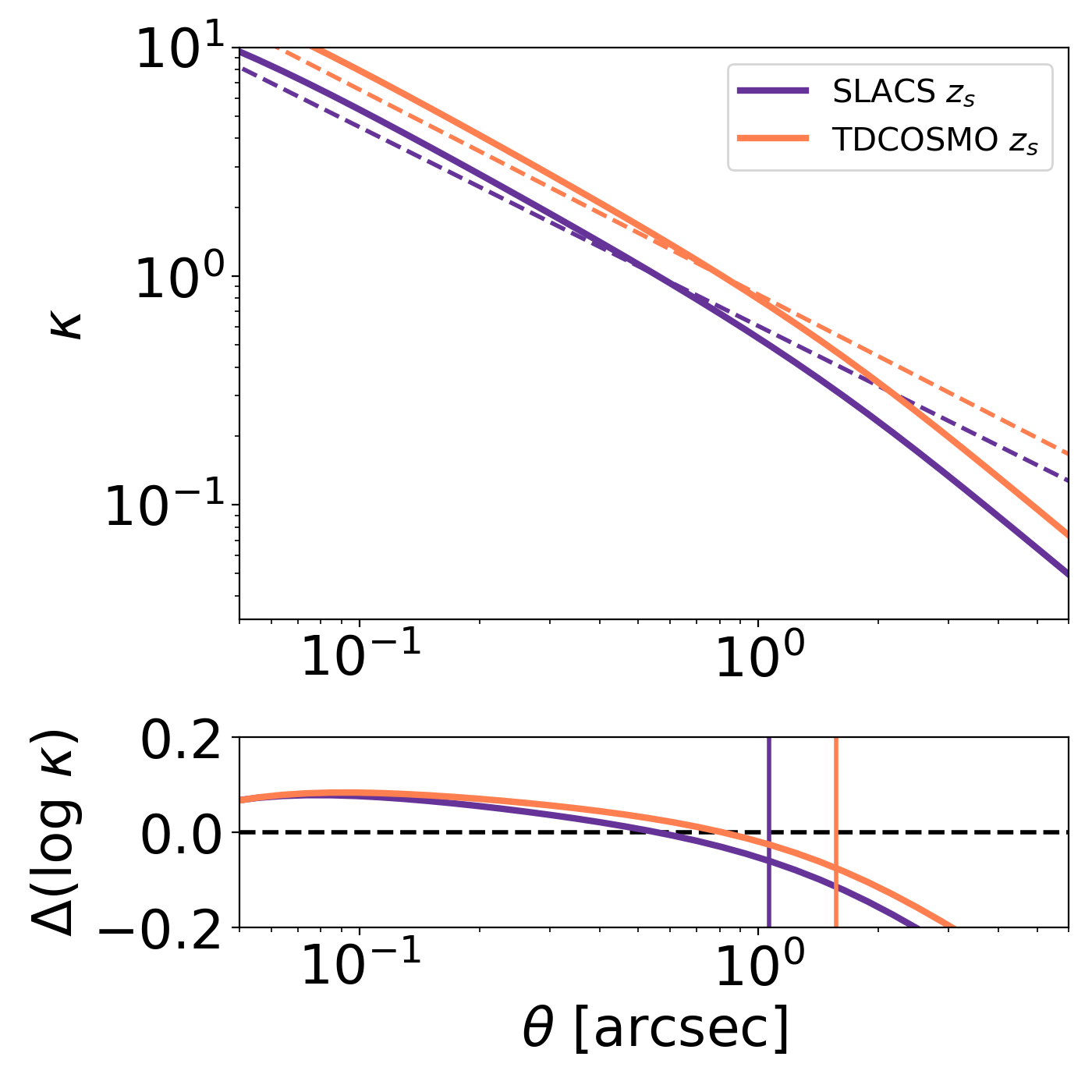}
    \caption{Example of a single profile, placed at each redshift and fit as a PEMD. Top: Convergence profiles as a function of radius; solid curves indicate input convergence profiles while dashed lines indicate convergence profiles corresponding to the PEMD fit. Bottom: residuals showing the differences between the input and PEMD fit profiles. Vertical lines correspond to the Einstein radii of the corresponding redshift.
    }
    \label{fig:kappa_pemdfit}
\end{figure}

As a demonstration of the MST, we plot the result from lensing fits to a single profile (created with a source placed at both $z_s$). Figure \ref{fig:kappa_pemdfit} shows that the best-fit PEMD result is not equivalent to the convergence profile, but rather has been transformed by an MST. This results in a shallower slope than the truth. Kinematics are required to correct the effect through a determination of $\lambda_{\rm int}$; the lensing data alone returns a biased value of $H_0$.

\section{Fitting parameter bounds}\label{sec:bounds}
The summarized parameters for lens construction and the uniform prior bounds used in this work are displayed in Table \ref{table:parambounds}.

\begin{table}
    \begin{tabular}{c | c | c}
        \centering
        
         Parameter & Value range & Notes \\
         \hline
         \hline
         \multicolumn{3}{c}{}\\
        \multicolumn{3}{c}{Lens mass distribution}\\
        \hline
         $q$ & [0.64, 0.78] & approximate, see Fig \ref{fig:param_properties}\\
         $\theta_{\rm eff}$ (\arcsec)& [1.43, 1.98] & approximate, see Fig \ref{fig:param_properties}\\
         $n$ & 4 & fixed\\
         $\Sigma_{\rm eff}$ ($10^{8}$ M$_\odot$ kpc$^{-2}$) & [4.1, 7.5] & approximate\\
         $r_s$ (kpc) & [20, 44] & approximate\\
         $\rho_0$ ($10^{6}$ M$_\odot$ kpc$^{-3}$) & [2.6, 15] & approximate\\
         
         \multicolumn{3}{c}{}\\
         \multicolumn{3}{c}{Mock image/data quality}\\
         \hline
         Source position & within caustic & Uniform\\
         Source $\theta_{\rm eff}$ (\arcsec) & 0.1 & fixed\\
         Source $n$ & 3 & fixed\\
         Noise & see text & fixed, see Sect. \ref{ssec:mockimages}\\
         Host source mag & 20.5 & fixed, unmagnified\\
         Point source mag* & 21 & fixed, unmagnified\\
         Kin. apertures (\arcsec)& (0.67, 1.33, 2.0) & fixed\\ 
         Kin. seeing (\arcsec)& 0.7 & fixed\\
         $a_{\rm ani}$ & 1 & fixed\\
         $v_{\rm disp}$ uncertainty & 10\% & fixed\\
         $r_{\rm eff}$ uncertainty & 10\% & fixed\\
         
        \multicolumn{3}{c}{}\\
        \multicolumn{3}{c}{Lens modeling allowed parameter range}\\
         \hline
         $\theta_{\rm Ein}$ (\arcsec) & [0, 100] & \\
         slope: $\gamma$ & [0, 100] & \\
         $q$ & [0, 1] & \\
         ellipse PA ($^\circ$) &[-90, 90] & \\
         $x_{\rm center}$, $y_{\rm center}$ (\arcsec)& [-100, 100] & \\
         shear: $\gamma_{\rm ext}$ & [0, 0.7] & \\
         shear PA ($^\circ$) &[-90, 90] & \\
         source $\theta_{\rm eff}$ (\arcsec)& [0,100] & \\
         source $n$ & [0.5,8] & \\
         source $x$, source $y$ (\arcsec)& [-100, 100] & \\
         
         \multicolumn{3}{c}{}\\
        \multicolumn{3}{c}{\texttt{hierArc} priors}\\
        
         \hline
         $H_0$ (km s$^{-1}$Mpc$^{-1}$) & [1, 150] & Uniform\\
         $\Omega_M$ & 0.3 & fixed\\
         $\lambda_{\rm int,0}$ & [0.5, 1.5] & Uniform\\
         $\sigma(\lambda_{\rm int})$ & [0, 1] & Uniform\\
         $\alpha_{\lambda}$ & [-1, 1] & Uniform\\
         $a_{\rm ani}$ & [0.1, 5] & Uniform\\
         $\sigma(a_{\rm ani})$ & [0, 5] & Uniform
    \end{tabular}
     \caption{Parameters for the mock creation and \texttt{hierArc} priors for this work. }
     \tablefoot{Where listed as "approximate", the approximate 16th and 84th percentiles are listed. *Only the TDCOSMO-like mocks have point sources}
     \label{table:parambounds}
     
\end{table}

\section{Single-redshift test}\label{sec:PSSLACS}

\begin{figure*}
    \centering
    \includegraphics[width=0.95\textwidth]{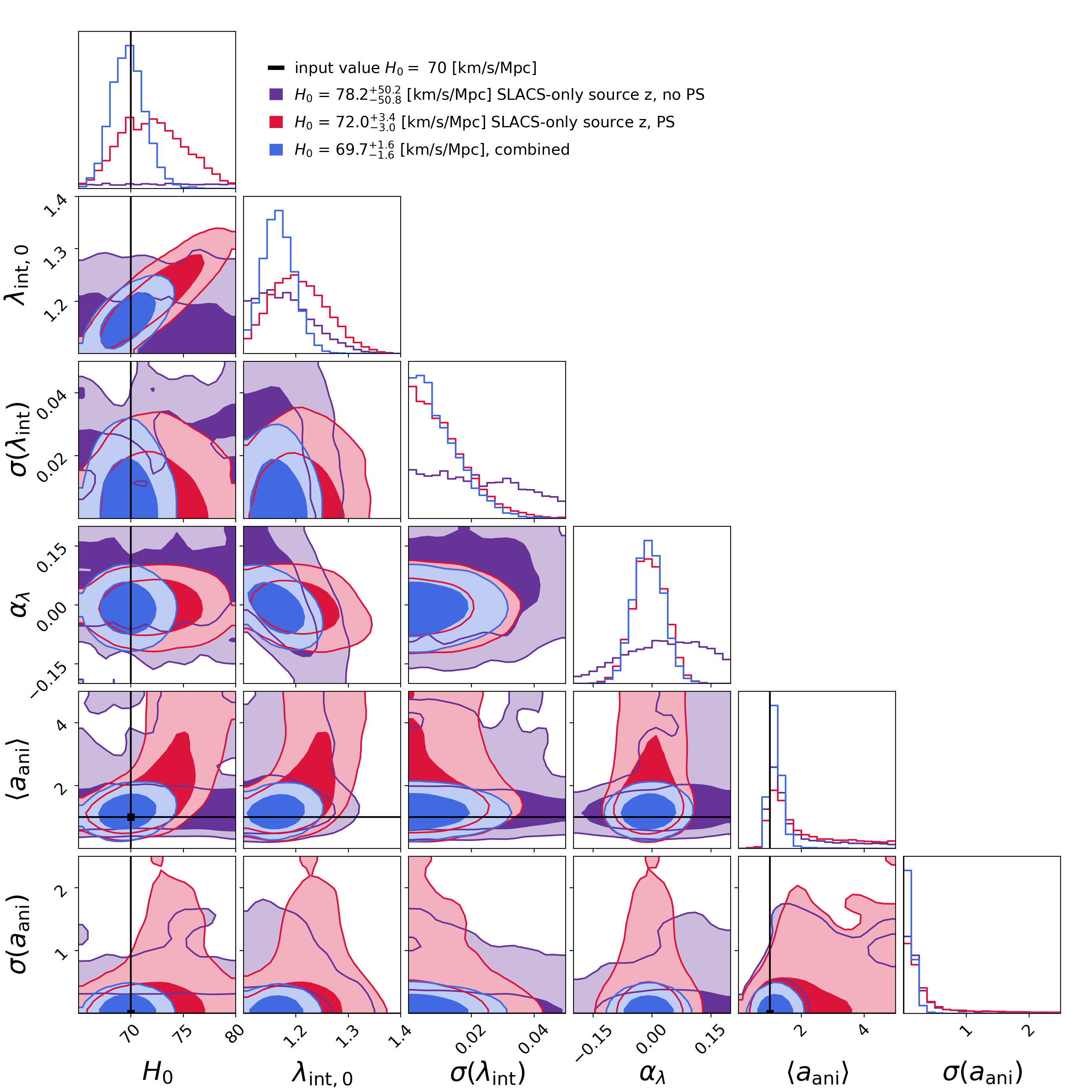}
    \caption{Hierarchical results from lens populations without point sources (purple) and with point sources (red) are plotted, as well as their combination (blue). Unlike Fig. \ref{fig:hierarc20pop}, the red population has the same source redshift as the purple population.
    }
    \label{fig:PS_SLACS}
\end{figure*}

Because the hierarchical framework is able to combine systems which span a range of impact parameters into an unbiased $H_0$ result, we were curious if the range of spread was beneficial to the success of the method. We modified the set of 20 SLACS-like images from Sect. \ref{ssec:20systems} by adding point sources with time delay information. For a set with no point sources, we use the remaining 20 SLACS-like images created in Sect. \ref{ssec:40systems}. As such, all lenses in this test have the same redshifts, with the case with no point sources simply corresponding to a different realization of profiles drawn from the same global distribution. We combine these sets together using the hierarchical framework and plot the results in Fig. \ref{fig:PS_SLACS}. 

We find that the result is similar to Sect. \ref{ssec:20systems}, with the combination of populations successfully recovering $H_0$ with $2.2\%$ precision. One minor difference is that the intrinsic scatter $\sigma(\lambda_{\rm int})$ is now consistent with zero; the framework captures the fact that this population of lenses has less intrinsic variation than the case where it is being probed at a wider range of different radii.

\end{document}